\documentclass[sigconf,balance=false]{acmart}

\usepackage{color}
\usepackage{graphicx}
\graphicspath{{figures/}}
\usepackage[labelformat=simple]{subcaption}
\usepackage{xspace}
\usepackage{multirow}
\usepackage{makecell} 
\usepackage{enumitem} 
\usepackage[ruled,vlined]{algorithm2e}

\usepackage{ulem}
\usepackage[nolist,nohyperlinks]{acronym}

\newcommand{\etal}{et~al.\xspace}
\newcommand{\eg}{e.g.\xspace}

\newcommand{\parhead}[1]{\medskip \noindent \textbf{#1}\hskip .1in}

\newcommand{\vsix}{IPv6\xspace}
\newcommand{\wifi}{Wi-Fi\xspace}
\newcommand{\ebay}{eBay\xspace}
\acrodefplural{WPS}[WPSes]{Wi-Fi Positioning Systems}

\ifdefined\isFinalized
\newcommand{\NewCommentType}[3]{}
\else
\newcommand{\NewCommentType}[3]{\expandafter\newcommand\csname #1\endcsname[1]{{\color{#2}{#3: ##1}} }}
\fi

\usepackage{popets}

\setcopyright{popets}
\copyrightyear{YYYY}

\acmYear{YYYY}
\acmVolume{YYYY}
\acmNumber{X}
\acmDOI{XXXXXXX.XXXXXXX}
\acmISBN{}
\acmConference{Proceedings on Privacy Enhancing Technologies}
\begin{document}

\title[Buy it Now, Track Me Later:\\Attacking User Privacy via
Wi-Fi AP Online Auctions]{Buy it Now, Track Me Later:\\Attacking User Privacy via
Wi-Fi AP Online Auctions}

\author{Steven Su}
\orcid{0009-0006-3409-6479}
\affiliation{%
  \institution{University of Maryland}
  \city{} 
  \state{} 
  \country{} 
}
\email{stevensu@umd.edu}

\author{Erik Rye}
\orcid{0000-0002-8151-8252}
\affiliation{%
  \institution{University of Maryland}
  \city{} 
  \state{} 
  \country{} 
}
\email{rye@umd.edu}

\author{Dave Levin}
\orcid{0000-0003-4957-5131}
\affiliation{%
  \institution{University of Maryland}
  \city{} 
  \state{} 
  \country{} 
}
\email{dml@umd.edu}

\author{Robert Beverly}
\orcid{0000-0002-5005-7350}
\affiliation{%
  \institution{San Diego State University}
  \city{} 
  \state{} 
  \country{} 
}
\email{rbeverly@sdsu.edu}

\begin{abstract}
  
Static and hard-coded layer-two network identifiers are well known to 
present security vulnerabilities and endanger user privacy.  In this
work, we introduce a new privacy attack against \wifi 
access points listed on secondhand marketplaces.  Specifically, we
demonstrate the ability to \emph{remotely}
gather a large quantity of 
layer-two \wifi identifiers by 
programmatically querying the eBay marketplace and 
applying
state-of-the-art computer vision techniques to extract IEEE 802.11
BSSIDs from the seller's posted images of the hardware.  
By leveraging data from a global Wi-Fi Positioning System (WPS)
that geolocates BSSIDs, we
obtain the physical locations of these devices both pre- and
post-sale.  In addition to validating the degree to which a seller's
location matches the location of the device, we examine cases of
device movement---once the device is sold and then subsequently
re-used in a new environment.  
Our work highlights a previously unrecognized privacy vulnerability
and suggests, yet again, the strong need to protect layer-two network
identifiers.

\end{abstract}


\maketitle

\section{Introduction}
\label{sec:intro}

Online auction websites like \ebay have been a popular way for users to
buy and sell goods online for  three decades.
Sellers post pictures and descriptions of their goods, which entice and
inform users who may choose to bid on them or directly buy them at a
fixed price.

Sellers and potential buyers both have reasonable expectations of
privacy.
Other than learning about one another's identities and addresses to
ship purchased goods, sellers expect that their public location
information is limited to what is explicitly shared on the auction
website (often just the city and state from which they are shipping).
Similarly, buyers can typically expect that nobody learns their
location other than the shipping address provided to the seller.
Certainly, all users expect that their locations before and after the
auction has taken place are not divulged to the public.

In this paper, we show that merely by examining the publicly available
photos on auction pages, we are able to successfully track the
locations of various network devices commonly sold on \ebay: \wifi \acp{AP}.
The key insight behind our work is that sellers often include pictures
of all sides of the devices they are selling, and \acp{AP} frequently
have their MAC address(es) written on them. 
Combining novel applications of computer vision to extract these MAC
addresses along with recent attacks that geolocate \acp{AP} from their
MAC addresses, our attack is able to track both sellers' and buyers'
locations in an automated and longitudinal manner.
In addition to traditional \acp{AP}, our techniques also apply to
devices that act as \acp{AP} as part of their device setup, such as
smart-home IoT devices.

Over approximately 5 months, we gathered 788k auctions inclusive
of \wifi \acp{AP}, \wifi-enabled home routers, and a variety of
\wifi-enabled IoT devices, \eg, smart electrical plugs.  Using modern
computer vision techniques, we process \wifi hardware advertisements
and extract \acp{BSSID}---the MAC addresses APs
announce---from the seller's images from 144k auctions.  

Then, by leveraging data from a global \ac{WPS} that geolocates
\acp{BSSID}, we obtain the physical locations of 13k BSSIDs from
devices both pre- and post-sale. 
In other words, in many scenarios, \emph{our attack exposes both the
buyer's and seller's locations}, long before and after the auction has
taken place.

We perform additional experiments and data analyses to validate the
degree to which a seller's location matches the location of the device.
We also present several case studies of the value and danger of being
able to track devices: we show, for instance, that some used network
hardware has been purchased and ultimately used at U.S.~military bases,
and that some devices marketed as ``new'' have in fact been %
previously used.

\parhead{Contributions}
We make the following primary contributions:
\begin{enumerate}
 \item Demonstration of the ability to gather 
   \wifi \acp{BSSID} unwittingly shared in online auction images
        programmatically and at-scale (\S\ref{sec:methodology}).
 \item Introduction of a novel attack that precisely geolocates 
   the gathered \acp{BSSID}, potentially compromising sellers' and buyers' 
        privacy (\S\ref{sec:results} \& \S\ref{sec:auction-geo}).
 \item Examination of several use cases, such as identifying secondhand
	 devices used on military sites, and identifying false claims of
		``new'' devices (\S\ref{sec:casestudies}).

 \item A proof-of-concept demonstration that this attack generalizes to other
     online marketplaces, such as Craigslist and Facebook Marketplace
        (\S\ref{sec:limitations}).
 \item Recommendations for attack remediation (\S\ref{sec:recommendations}).
\end{enumerate}

\section{Background and Related Work}
\label{sec:background}

\subsection{\ebay and Online Auctions} %

\ebay is an online auction website and marketplace founded in 1995. \ebay
allows users to sell new or used items in second-price ascending auctions, or to
sell them for a set price (called the ``Buy It Now'' option). When a buyer wins
an auction or purchases an item directly, their identity and address are
disclosed to the seller by \ebay for the purpose of shipping the purchased
goods. \ebay does not disclose the buyer's identity to any other user of the
website.  Sellers' registered identities and the city and state from which they
will ship their items are made public on each auction item page so that buyers
can make more informed decisions as to their trustworthiness and shipping
times.

Several other prior studies have used online marketplaces to glean sensitive
information about users.
Minkus and Ross studied the \ebay feedback system, which allows buyers and
sellers to rate the quality of their transactions with each
other~\cite{minkus2014know}. These ratings alert future buyers and sellers to
the trustworthiness of the individual they are transacting with. However, the
implementation of \ebay's feedback system divulges information about the buyer's
purchase history, including potentially privacy-sensitive items they have
purchased. In contrast, we use the photographs users post online, rather than
attempting to correlate user activity over time. Additionally, rather than learn
about their purchase history, our attack exposes users' precise locations.

Several studies have demonstrated that sensitive information can be obtained by
purchasing goods on public auction.
Roberts~\etal purchased hundreds of cell phones from a police auction website
to determine what types of privacy protections were implemented for the users
whose (former) property was for sale~\cite{roberts2023blue}. They found that in
many cases, the users' data was still present on the device and accessible due
to poor screen lock passcodes and patterns. Garfinkel~\etal purchased hard
drives on secondary markets (including \ebay) to analyze disk sanitization
practices~\cite{garfinkel2003remembrance}, finding that roughly a third of hard drives
they purchased had recoverable sensitive information.

To the best of our knowledge, our work is the first to show that it is possible
to learn private location information about both buyers and sellers without
purchasing any items, based only on the pictures made publicly available on the
auction website.

\subsection{MAC Addresses \& \wifi Router Geolocation} %

One of the central ideas behind our attack is to extract \ac{MAC} addresses from
public auction photos, and use them as a persistent, trackable identifier.
We review the relevant details of MAC addresses and MAC address-based tracking here.

MAC addresses are 48-bit identifiers used to indicate the source and destination
of link-layer frames; they are used in 802.3 Ethernet, 802.11 \wifi, and
Bluetooth. MAC addresses are typically written as 12 hexadecimal characters, with
each pair separated by a colon or hyphen, \eg, \texttt{a0:2b:ca:92:1c:da} or
\texttt{10-29-ca-2a-be-2f}. MAC address space is managed by the IEEE, which assigns
blocks of $2^{24}$ MAC addresses in three-byte prefixes called
\acp{OUI} or MAC Address-Large (MA-L) allocations\footnote{Smaller allocations
are also offered, called MAC Address-Medium (MA-M) and MAC Address-Small (MA-S)
that contain $2^{20}$ and $2^{12}$ MAC addresses, respectively.}. The IEEE
publishes and regularly updates a list of assigned \acp{OUI}~\cite{oui}. The
privacy risks of static MAC addresses are well-known and studied, particularly
in 802.11
\wifi
clients~\cite{abedi2013bluetooth,goodin2013no,cunche2014know,freudiger2015talkative,matte2016defeating,vanhoef2016mac,martin2017study,martin2019handoff,becker2019tracking,celosia2020discontinued,uras2022mac,fenske2021three}.
Our work diverges from prior art by developing a novel way to learn \wifi MAC
addresses---by recovering them from photographs from online auctions and
marketplaces. This technique eliminates the requirement for an attacker to be
physically proximate to a device to learn its \wifi MAC address.
 
Network operators commonly use MAC addresses to identify and manage devices on
their network.
For this reason, it is common practice for manufacturers of networking
hardware---especially \wifi \acp{AP}---to print the device's MAC address(es) on
the device itself.
This facilitates quickly identifying devices, but as we show in this paper, it
also facilitates tracking some devices' precise locations and movements.

A \wifi \ac{AP} and the devices connected to a network it advertises form a
\ac{BSS}; the \ac{MAC} address the \ac{AP} uses for that \wifi network is known
as the \acf{BSSID}. Mobile client devices will frequently use random MAC
addresses while probing for nearby \wifi networks and after connecting to
them~\cite{vanhoef2016mac,martin2017study,matte2016defeating,fenske2021three,
freudiger2015talkative,cunche2014know,goodin2013no} for privacy reasons.
However, APs' MAC addresses typically do not change, because, until
recently~\cite{rye2024surveilling},
they were not considered as privacy-sensitive as client MAC addresses (in
large part because most APs are generally stationary for long periods).

Our work relies on geolocation information obtained from a \acf{WPS}.  \acp{WPS}
are systems run by operating system vendors and others to enable devices in
their ecosystems to self-geolocate without using GPS. Devices do this by
scanning for nearby \ac{AP} \acp{BSSID}, which they report back to the \ac{WPS}
operator. These nearby BSSIDs act as landmarks to help determine their location. 

Several works of prior art have geolocated \wifi \acp{AP} using \acp{WPS}.
Rye and Beverly carried out an active measurement campaign to obtain \vsix
addresses from home routers~\cite{ipvseeyou}. A subset of these embed their MAC
address in the lower 64 bits of the \vsix address, which they then geolocated
using Apple's \ac{WPS}, the open-source project WiGLE~\cite{wigle}, and other
open-source BSSID geolocation databases. Unlike Rye and Beverly's work, our
work requires no active measurements or specific types of \vsix addresses, but
rather derives \acp{BSSID} from photographs users post of their \acp{AP} online.

Rye and Levin used Apple's \ac{WPS} to conduct a longitudinal measurement of
\wifi \acp{AP}, which revealed that many move over
time~\cite{rye2024surveilling}. They used their large-scale dataset to detect
the positions and movements of troops in and around Ukraine, and destruction
and power outages in Gaza. In contrast, we conduct a more targeted privacy
attack in this work in which the BSSID of the target is known through its
disclosure in photographs posted online.

\begin{figure*}[t]
        \centering
        \includegraphics[width=\linewidth]{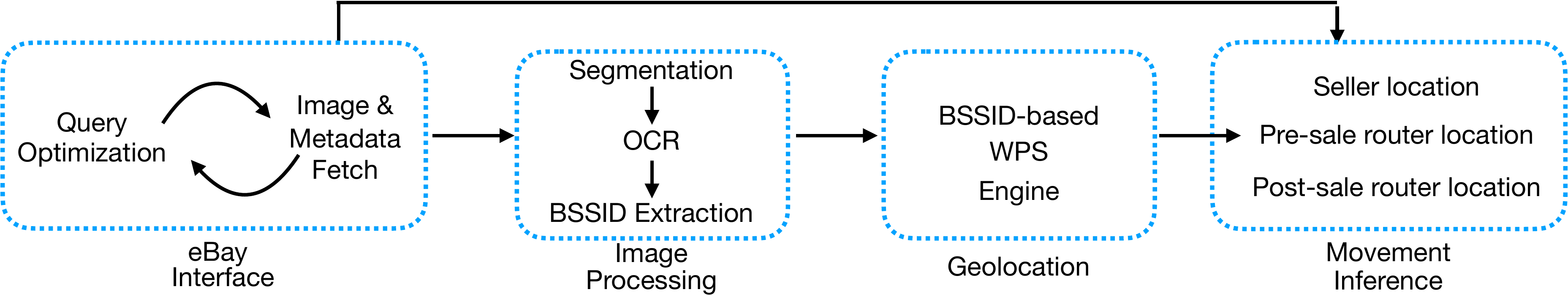}
        \caption{Methodology Overview:
          Images of \wifi hardware on eBay are processed to find and extract
          BSSIDs.  These BSSID are fused with precision WPS geolocation to
          create a novel privacy attack.}
        \label{fig:ebay}
\end{figure*}

Our work also explores the extent to which MAC addresses reveal devices'
specific models.
In 2016, Martin~\etal developed a novel method for determining the MAC address
allocation schemes by a variety of manufacturers with
model-granularity~\cite{martin2016decomposition}. To do this, they captured more
than 2B 802.11 frames while wardriving, and extracted model-identifying
information from \wifi Protected Setup information elements in 802.11 management
frames and \ac{mDNS} responses. These techniques relied on devices using
their global, hardware MAC address in order for the MAC address-to-model
mappings ranges to have any meaning; today, client devices using modern iOS and
Android operating systems use random MAC addresses before and after association
to prevent long-term tracking~\cite{fenske2021three}. Further, \wifi Protected
Setup suffers from numerous security and privacy
vulnerabilities~\cite{wps-brute,bongard2014offline,mohtadi2015new,costantin2017vulnerabilities,sanatinia2013wireless},
and \wifi Protected Setup information elements are now infrequently used by both
clients and access points.
In this work, we demonstrate that \ebay
photographs can be used to derive manufacturer model ranges too, without needing
to ever be within transmission range of a device in question.

\section{Methodology}
\label{sec:methodology}

To gather \acp{BSSID} and track locations of \wifi \acp{AP} listed in
eBay auctions, we develop a multi-stage pipeline.
The pipeline
ingests eBay auction data via an eBay-supplied API, downloads 
images of items for sale,
and uses \ac{OCR} to extract addresses (Figure~\ref{fig:ebay}). 
We discuss the
major components of this pipeline in the following sections.

\subsection{eBay API}
\label{sec:ebayAPI}

Unlike other popular online marketplaces such as Craigslist or Facebook Marketplace, 
\ebay provides an official and documented API for developers to create 
integrations that augment the experience of buyers and sellers. 
The API provides a search endpoint intended to aid product discovery in software integrations
and permits
keyword queries, 
filters, and other search refinements \cite{ebay-search-documentation}. 
Product listing query results returned via the API include a variety of information, 
including the title, links to images provided by the seller, price, condition, and seller among other data. Search results can be further expanded 
and refined by a 
``fieldgroups'' API parameter to obtain the seller's
coarse physical location, buying options 
(\eg auction, ``Buy it Now''), filters by brand, 
and gather item details specific to wireless network
hardware, \eg,
its highest implemented \wifi standard,
data rates supported, etc.

\ebay images embed an implied ID, image size and format in their
URLs within the path
\texttt{/images/g/<ID>/s-l<size>.<format>} on 
the \texttt{i.ebayimg.com} domain.
Manual experimentation varying the size and format 
allow us to obtain images between 32 to 2400 pixels horizontally and in
JPEG, PNG, or WEBP formats.
To balance quality and storage, we request 1600 pixel files in JPEG
format.
All available images for a given listing were downloaded and processed separately in our pipeline.

Auction listings use an identifier that is
unique across all marketplaces and allows us to 
de-duplicate collected data. This identifier is returned both by the API
and used as the listing URL \\
(\texttt{www.ebay.com/itm/<ID>?<HTTP GET args>}).

\subsubsection{Exhaustive API Search}

One notable limitation of the \ebay API is a maximum paginated results set of
10,000 results. This limitation also appears in \ebay's website search---while a query may report a significantly
larger results set (e.g., 71,000+ results for a query of ``router'' on the
United States
eBay marketplace), a user can only view a maximum of 10,000 results
before artifacts of the limitation start to appear,
such as repetition of listings already shown in previous pages.
Experimentation with the API reveals that \ebay's 10,000 results limit applies to a given query
combined with any applied filters, such as the product's brand or features. 
Thus, by providing unique combinations of filters in addition to the search terms of a query, 
we can effectively obtain more than the 10,000 results limit for a given search query text alone.

Therefore, by dividing the potential results for a given query by utilizing sets of mutually exclusive filters,
we can efficiently and potentially exhaustively enumerate all possible results for a given query. In our system, for a given query,
we first send a ``probe'' request to eBay's API to obtain a listing of possible
options for the ``brand'' filter and the number of listings associated with each filter option. 
Next, 
to best utilize limited API calls, we use a bin packing algorithm to group filter options such that the combined listings available for each group fill the maximum 10,000 results 
(or as close as possible). Then, for each group of brand filter options, we generate sets of API calls that exhaustively query all listings via the API's pagination mechanism and store the data.
We refer to this technique in future sections as a full or exhaustive query.

\subsubsection{International Marketplaces}
While \ebay's primary \texttt{ebay.com} domain serves its United States marketplace, the platform currently operates in a few dozen 
international marketplaces. The \ebay API allows developers to specify
the marketplace ID.
To understand how our findings might generalize beyond the U.S. marketplace, we query
a subset of these marketplaces 
by translating search terms and filters to the marketplace's primary
language.

We observe diminishing returns in the number of unique auctions as we included additional marketplaces. 
First, some countries appear to have smaller second-hand markets on
\ebay for \wifi APs and related devices. 
Second, it appears that listings are not mutually exclusive between marketplaces. \ebay appears to cross-list product 
listings from nearby marketplaces, (e.g., Canada and United States marketplaces in North America) 
especially if the listing supports international shipping. 
Considering these factors and \ebay's daily API call quota, we generally only 
collect \ebay listings from the larger marketplaces, such as marketplaces located on their own country's top-level
domain rather than a subdomain of \ebay's primary .com site or those having a
reasonable number of listings for \wifi \acp{AP} and related 
products after manual testing. 

\subsubsection{Search Queries}
For each marketplace, we utilize two sets of search queries. Per \ebay's documentation, queries 
are case-insensitive and most punctuation marks are ignored when
matching keywords (i.e., no differentiation between \texttt{wifi} 
and \texttt{Wi-Fi}). The first set of queries are broad but still reasonably 
targeted terms to find products likely to contain \ac{MAC} addresses in the listing images. The specific two queries 
in this first set, as translated for English-speaking marketplaces, were \texttt{wifi router} and \texttt{wifi access point}.
This first set of ``general'' queries, \texttt{\{wifi router, wifi access point\}} was used in their language-translated form in all marketplaces.

The second set of search queries consists of brand-specific terms to
retrieve listings for \wifi  devices manufactured
by specific companies. The choice of brands to target was determined by 1) popularity of the brand in a given marketplace as
indicated by the counts provided by eBay for the appropriate brand filter or 2) our observation of products that would be expected to appear
in \ebay's search results using the general queries but for unknown reasons do not (e.g., Verizon ISP routers). These targeted queries are 
especially useful to obtain listings where the seller did not include identifying terms such as \texttt{router} by instead only used the brand 
and model number in the title or cases where certain brands simply did not appear in search results of the more general queries due to
idiosyncrasies in eBay's search algorithms. The design of this second set of queries utilizes \ebay's advanced search \cite{ebay-advanced-search-documentation} to more
narrowly restrict results. 
The advanced query language allows clauses in the format \texttt{(term1, term2,
term3, \dots)}. Each of the search terms 
functions as a logical \texttt{OR} in the search, meaning listing titles must contain at least one of the comma-separated terms in the clause
to be shown in the search results. Similarly, one may add a dash in front of the clause (i.e., \texttt{-(term1, term2, term3, \dots)})
to create an exclusion list where the presence of one or more terms stops a listing from being returned.

Furthermore, when specific query filters are applied, 
\ebay's search behavior changes. Specifically, \ebay provides ``automatic keyword expansion'' to simple queries that do not utilize 
advanced query language. Per their documentation, automatic keyword expansion results in a wider variety of listings and is more forgiving 
in determining matching listings by, for example, also matching on synonym terms. Therefore, our usage of these advanced filters requires 
consideration on a balance of breadth of relevant search results and inclusion of irrelevant, ``noisy'' results. 
It is important to note that noisy results do not cause false
inferences or errors in our pipeline (we simply do not find any
relevant MAC addresses).  However, we seek to minimize this noise in
order to optimize the pipeline and limit our query volume.

With all of the above in consideration, the second set of search queries are those that follow the format: 
\texttt{\{brand\}} \texttt{(<inclusion terms>)} \texttt{-(<exclusion terms>)}.
The specific inclusion terms used were \texttt{ap}, \texttt{access point},
\texttt{router}, \texttt{radio}, \texttt{wifi}, \texttt{wireless}, and
\texttt{mesh}, while the specific exclusion terms were \texttt{car},
\texttt{carplay}, \texttt{mouse}, \texttt{phone}, and {gb}. One query following
this 
format was generated for each brand targeted in each marketplace. The specific brands targeted are not exhaustively listed
here for brevity, but include common consumer brands such as Ubiquiti and Netgear, enterprise-grade brands such as Ruckus and Aruba,
and ISPs such as Verizon. The second sets of brand-specific queries were used only in English-speaking marketplaces (i.e., US, GB, CA, AU)
with the brands tuned for each specific marketplace as previously described.

\subsection{Query Schedule}
A primary limitation is \ebay's quota of 5,000 search API calls per day for a given credential set \cite{ebay-api-limit-documentation}. 
While it is possible to request additional API key pairs or to 
utilize multiple accounts, we aim to ensure efficient usage of API calls to reduce impact on eBay's servers. 
All marketplaces are queried fully twice a day at staggered 12 hour intervals
with marketplaces grouped to spread out the total volume of queries in time.
We perform full and exhaustive searches 
for both query sets (i.e., the general and brand-specific queries)
relevant to each marketplace.

Unfortunately, we were not able to incrementally query for new results
due to limitations in eBay's interface.
eBay's search function and API do not provide a direct sort for listing time; only rough analogues 
such as ``Newly Listed'' and ``Ending Soonest'' listings are available. 
Manual testing reveals that these sort options do not behave as a na\"ive sort by time as the reported size of the results set in 
the site's web interface differs depending on the sort option. Nonetheless, we find that sorting by ``Newly Listed'' still 
appears to surface more recent listings and useful results.

\begin{figure*}[t]
        \centering
        \includegraphics[width=0.9\linewidth]{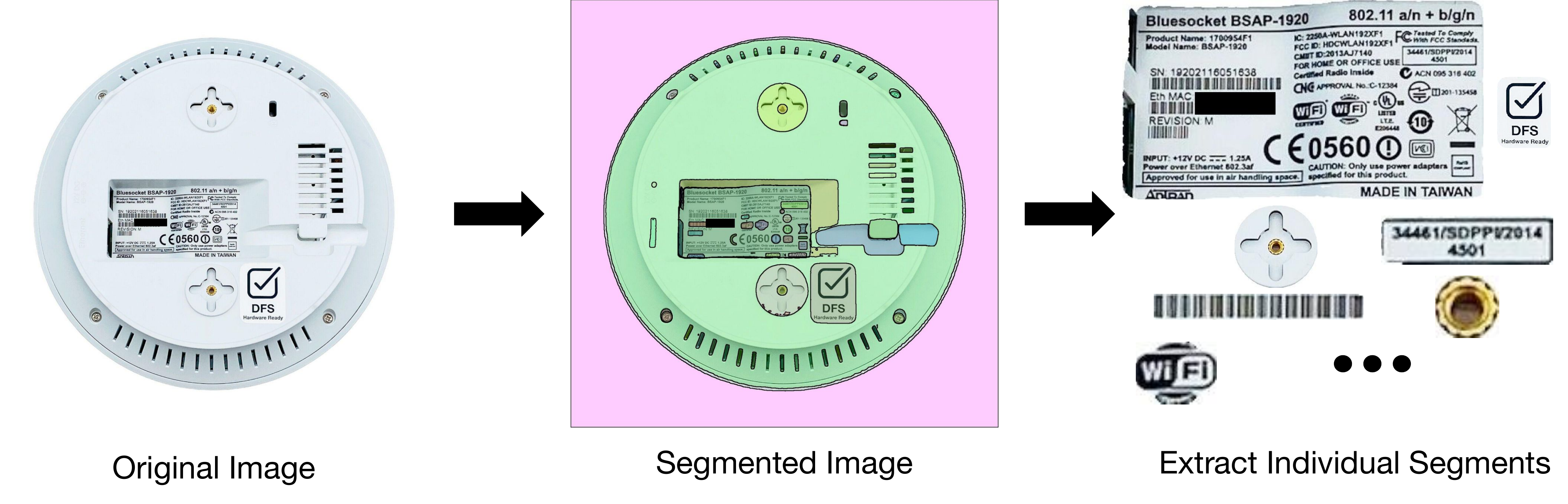}
        \caption{Segment Anything model segmentation masks and segment extraction}
        \label{fig:segmentation}
\end{figure*}

Thus, we developed a second search methodology different from the full and exhaustive search. Compared to the
methodology described for the full and exhaustive search, we utilize the ``Newly Listed'' sort option instead
of the default ``Best Match''. Additionally, rather than paginating to the maximum 10,000 results for a given
query and filter set, we halt pagination of results when the timestamp of the next listing exceeds the timestamp 
of the last stored listing in our database prior to initiation of the API querying sequence. Finally, due to the
nature of this search method, we omit the initial probe request and partitioning by brands. The lightweight nature
of this search method means that parallel to the scheduled twice-a-day full and exhaustive searches for each group
of marketplaces, we run this second search methodology at three hour intervals for every query and marketplace
combination without regard for the marketplace groupings used in the full and exhaustive searches.

\subsection{\ac{OCR} \ac{MAC} Address Extraction}
\label{sec:ocrmethodology}

In this section, we describe the machine learning components of our pipeline used to automate extraction of MAC
addresses from the collected eBay listing images and the insights that informed
the choice of models. Our
pipeline consists broadly of three stages: 1) image segmentation to separate an image into broad regions of
 interest, 2) optical character recognition (OCR) to extract all text from image segments, and 3) text processing 
to isolate potential MAC addresses from extracted text.

\subsubsection{OCR Model Selection}

The final models selected for the pipeline were chosen primarily based on observed accuracy on a small, manually labeled
dataset of images and corresponding MAC addresses while also considering throughput and cost. In the initial development
of the pipeline, approximately 80 images were chosen from eBay listings of Wi-Fi
\acp{AP}, curated such that all 
images had at least one MAC address present but with varying perspectives and degrees of text clarity.

We evaluated a number of popular and well-supported \ac{OCR} libraries
and models, as well as multi-modal large language models (LLMs)
with
zero-shot transfer capabilities.
During the testing process, some libraries were eliminated from consideration due to bugs and difficult-to-resolve requirements and dependency
issues. Others failed to include a complete pipeline and were also
eliminated. Final testing revealed the PaddleOCR library \cite{paddleOCR} and an 8-bit quantization of the Qwen2-VL-2B-Instruct model 
of the Qwen family of LLMs \cite{qwenvl} to be the best performing in accuracy
while also balancing compute requirements on available hardware. %
For the Qwen model, we observed an 68.75\% accuracy on the test set for finding at least one ground truth \ac{MAC} address in each image 
compared to the PaddleOCR model which achieved a 57.5\% accuracy for the same
criteria. The Qwen model, however, was significantly slower in inference time when 
implemented with the Python transformers \cite{wolf-etal-2020-transformers} library.
Therefore, PaddleOCR was chosen as a reliable and performant library both in accuracy and inference time sufficient for use in our pipeline.

\subsubsection{Pipeline Optimizations}
Having selected an OCR model, we investigated additional augmentations that could boost
OCR accuracy. Commonly used pre-processing techniques include background removal, binarization, 
and the application of filters to improve OCR performance. The below methods
were primarily tested against the PaddleOCR library, though we did implement
some augmentations with other models to confirm that PaddleOCR remained an
optimal choice.
We did not exhaustively explore combinations of models and optimization methods
to reduce search space of potential pipelines.

We manually tested multiple image binarization algorithms, sharpness and blur filters, denoising algorithms, and other image processing 
techniques with associated algorithm hyperparameters. No tested combination of techniques and hyperparameters appeared to significantly improve 
the performance of PaddleOCR; in some cases, accuracy was reduced. We eventually opted out of applying any of these pre-processing techniques
due to lack of impact and unknown appropriateness of filters given the wide variety of conditions for the full set of listing images (e.g., 
perspective, lighting, etc.).

\begin{figure*}[t]
        \centering
        \includegraphics[width=0.9\linewidth]{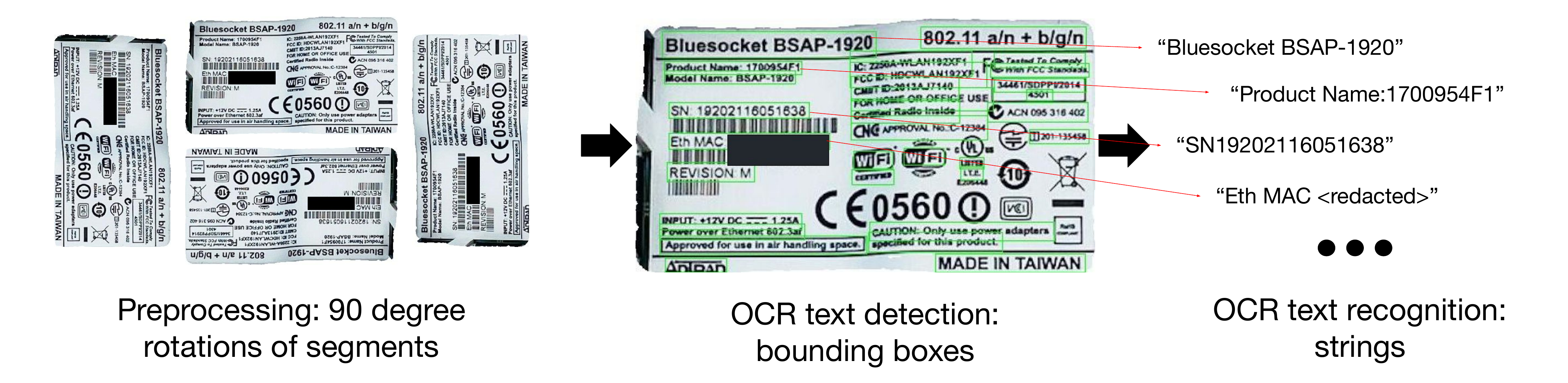}
        \caption{PaddleOCR image processing stages.}
        \label{fig:ocr1}
\end{figure*}

Another method explored was the usage of image segmentation to extract areas of an image likely to contain text. From 
manual exploration, it was apparent that most MAC addresses on router devices are printed on sticker or labels, usually
accompanied by other information such as default passwords, serial numbers, barcodes, and other device information. Processing 
only these regions with an OCR system could potentially improve accuracy by omitting extraneous background noise. Traditionally,
image segmentation requires a task-specific model and we were unable to find any widely-utilized models appropriate for extracting
objects such as labels from images. However, given the popularity of foundational
models, we found Meta's Segment Anything~\cite{kirillov2023segany} series of
models to be an interesting option. These models, which are available in a variety of sizes (i.e., number of parameters),
generalize the task of image segmentation to a more task-independent general case via a promptable segmentation system with zero-shot capabilities.
Testing the model on a sample of images revealed strong capabilities to extract labels containing MAC addresses. The behavior and output of this 
segmentation system is shown in Figure \ref{fig:segmentation}. 

Using Segment Anything to produce inputs for PaddleOCR resulted in improved MAC
address extraction accuracy. Further testing informed the choice of the large,
first generation Segment Anything model from an array of model sizes across two
generations. In our testing, this model best balanced accuracy and throughput.

We note that the image segmentation stage adds a significant computational
burden to the pipeline for an increase in accuracy. Besides 
the overhead of the model itself, by nature, the segmentation model also returns other segments in the image that are irrelevant to our purposes,
such as backgrounds and various physical components of an \ac{AP}. We are unable to filter out segments reliably without developing a classification model,
so all segments are processed through the OCR model adding additional compute time. One other optimization method we found useful was attempting OCR on
all 90 degree rotations of each segment. While the PaddleOCR library claims to be able to handle rotated text, our empirical testing revealed accuracy improvements
when applying this technique. The stages of the PaddleOCR library with the 90 degree rotations on an extracted segment are shown in Figure \ref{fig:ocr1}.
In order to process the backlog of historical eBay listings as well as keep up with the ingest rate of new daily listings, we utilized GPU
compute nodes our institution's cluster. However, a lower resource attacker could potentially obtain acceptable results at higher throughput by forgoing
the additional optimization steps we added to our pipeline. Overall, when combining the segmentation and rotation techniques with PaddleOCR as the \ac{OCR} library,
we observed an 82.5\% accuracy on the test set for finding at least one ground truth \ac{MAC} address in each image and 77.5\% for finding all \ac{MAC} addresses
present in each image. We believed the throughput and accuracy of this pipeline to be sufficient to proceed.

\subsubsection{Regular Expression Extraction}
The output of the OCR pipeline is lines of text and their associated positions
in the input image. To extract MAC addresses from this output,
we combine all text together and normalize by removing all non-alphanumeric characters and converting to lowercase. We apply a simple regular expression
matching the format of a MAC address (with or without separators between octets) to find MAC address candidates. The extraction process is intentionally likely
to produce false positives, matching on values such as serial numbers, as the impact of false positives is minimal when querying the WPS database. Candidate
MAC addresses are further filtered for addresses whose first three octets form a
valid, IEEE-assigned OUI~\cite{oui}.

\subsection{\ac{MAC} Address Extraction Validation}
\label{sec:macvalidation}

To validate the results of our pipeline and understand the performance of the \ac{MAC} address extraction,
we manually annotated a small subset of images and compared the human-extracted addresses with those identified
by the system. To create the annotation subset, all images that were processed by the system were partitioned
into three mutually exclusive groups: 1) images that had one or more \ac{MAC} addresses (with or without valid \ac{OUI}s) identified by the pipeline,
2) images without any \ac{MAC} addresses extracted but having at most ten words extracted by \ac{OCR}, and 3) images without any \ac{MAC}
addresses extracted and having more than ten words identified. The partitions represent 8.8, 33.8, and 57.4\% of the entire
dataset of images, respectively. For the first partition, we allow images whose extracted \ac{MAC} addresses 
have invalid \ac{OUI}s to account for cases where \ac{OCR} errors result in the
recognition but incorrect transcription of
\ac{MAC} addresses. The purpose of the latter two groups is to identify cases
where a MAC address is likely present but not identified by the system. It appeared that only a relatively small percentage 
of images actually contain a \ac{MAC} address relative to images that do not and without any intervention, we believed 
it would be difficult to identify failure cases. The logic behind the partition is that we generally expect images of 
the backs of \acp{AP} to have few words while images with many words are expected to be other items such as instruction manuals 
or the backs of boxes with product information.

A random sample of 250 images were extracted from each partition. The collective 750 images to be annotated were then
presented in mixed but consistent order to multiple trained reviewers who were asked to independently identify and transcribe all \ac{MAC} 
addresses present in the image. Input was performed on a custom-designed web interface with basic validation on the format
of \ac{MAC} addresses. Reviewers were asked to note special cases such as the presence of \ac{MAC} address redaction. In the case of significant
uncertainty, reviewers were asked to make their best guess at individual characters or use placeholders in the worst case scenario but to
nonetheless record a value indicating presence of a \ac{MAC} address.

In determining ground truth \ac{MAC} address annotations for each image, for consistency and internal validation, we 
require annotations from two reviewers and only \ac{MAC} addresses that show agreement by both reviewers are accepted as ground truths; all
other annotated \ac{MAC} addresses are rejected. In cases where multiple reviewers annotated an image, the 
annotations from the first two reviewers ordered by internal reviewer ID were
selected. Overall, 45 of the 750 images were considered inconclusive after annotation where the pipeline identified \ac{MAC}
addresses but no addresses with sufficient agreement were found in annotations.
Eighty-five percent of rejected \ac{MAC} addresses did not match any
addresses identified in the pipeline. The cases of rejected \ac{MAC} addresses likely result from annotator data entry errors causing
lack of agreement or ambiguous cases (e.g., blurry images or partial obstructions of \ac{MAC} addresses) where human annotators are
able to recognize the presence of but not agreement on the actual \ac{MAC} address and the pipeline may or may not be able to identify
an address at all depending on image quality.

\begin{table}[t]
    \caption{Performance extracting MAC addresses from eBay images,
using human annotators as ground truth on three different image
partitions.}
    \label{tab:annotation_counts}
\begin{tabular}{r|c|c|c|c}
& \multicolumn{1}{c}{\textbf{Partition 1}} & \multicolumn{1}{c}{\textbf{Partition 2}} & \multicolumn{1}{c}{\textbf{Partition 3}}& \multicolumn{1}{c}{\textbf{Total}} \\
\hline
\textbf{TP} & 150 & 0 & 0 & 150 \\
\textbf{FP} & 60 & 0 & 0 & 60 \\
\textbf{FN} & 22 & 2 & 2 & 26 \\
\textbf{TN} & 63 & 245 & 238 & 546
\end{tabular}
\end{table}

For each image, unique \ac{MAC} addresses from the pipeline and annotations are classified into four categories. True positives (TP) are counted
each time an accepted ground truth matches an address identified from the pipeline. False positives (FP) are counted for each \ac{MAC} address 
identified by the pipeline that does not match an accepted ground truth from annotation. Similarly, false negatives (FN) are counted for each
accepted ground truth not identified by the pipeline. Finally, a true negative (TN) is counted for an image that had no accepted ground truths
(but was not inconclusive as previously discussed) and no \ac{MAC} addresses as identified by the system. Only filtered \ac{MAC} addresses
(those with valid \acp{OUI}) were considered from the pipeline for
classification purposes. The overall counts are presented in Table~\ref{tab:annotation_counts}.
Count totals are higher than the number of images in each partition since each
image may have more than one unique \ac{MAC} address from the pipeline and annotations
and each address (or lack thereof) contributes to a category.

The partition 1 counts show a few interesting artifacts from the design of the pipeline. The relatively high true negative count of 63, which may be unexpected 
for a partition of images supposedly containing \ac{MAC} addresses, represent images where the pipeline extracted text resembling \ac{MAC} addresses and hence
met the requirements for inclusion in the first partition, but all addresses were filtered out for invalid \ac{OUI}s and therefore counted as true negatives with 
the corresponding empty annotations. At the same time, the false positive count of 60 indicates there are many cases where even filtering for invalid \ac{OUI}s was insufficient due to \ac{OCR} errors in other locations
of a \ac{MAC} address or that the image quality resulted in sufficient ambiguity that annotators could not agree whether a pipeline extracted address is correct. The false negative count of 22 is similarly
indicative of cases where \ac{OCR} errors resulted in the failure to identify annotator agreed-upon addresses. Partition 1 had the highest number of inconclusive images (12.8\% of images in the partition),
further showing that a non-trivial portion of images containing \ac{MAC} addresses were of such poor quality that even human annotators are unable to agree on ground truth addresses.

The counts for partitions 2 and 3 fell within expectations. The majority of results were true negatives which makes sense given these two partitions
did not have any \ac{MAC} addresses extracted by the system in the first place. Two false negative cases were situations where the image was too blurry
to extract \ac{MAC} addresses and human annotators were only able to recognize the presence of \ac{MAC} addresses from experience but could
not reasonable identify the actual characters and left placeholders. The other two false negative cases were images with extreme camera angles that
human annotators could read but were missed by the pipeline. These two partitions had 13 images that were inconclusive and mostly consisted of images
with \ac{MAC} address but poor image quality made it difficult for human annotators to agree on ground truth values.

Overall, aggregating across all three partitions, the pipeline exhibited
relatively high accuracy of 89\% in identifying correct \ac{MAC} addresses in
images with and
without \ac{MAC} addresses. If the counts in each partition are weighted by the partition's representation of the entire dataset, the accuracy rises to 96.3\%. As intended, 
the overall precision of 71\% and false positive rate of 10\% show the effects of permissive, regular expression-based \ac{MAC}
address extraction from \ac{OCR} outputs with the help of \ac{OUI} validation. But, as discussed previously, the impact of extracting \ac{MAC} addresses that are in reality
invalid are minimal as they would simply not show up in a \ac{WPS}. We believe that these overall statistics show that our system is able to successfully extract
\ac{MAC} addresses from listing images when present with reasonable confidence in the output.

\section{Ethical Considerations}
\label{sec:ethics}

Our institution's IRB reviewed our experiment design and determined that it is
not human-subjects research.
According to our IRB office, this determination was based on three
observations:
1)~Our experiments involved no interaction or intervention with
human subjects,
2)~All of the data we collected was public information, and
3)~We did not collect information about any humans themselves, but rather
about devices, and thus we were not collecting personally identifiable
information.
While one could certainly use the device location information that we obtain to
infer information about individuals (e.g., by looking up the address where
the device is located and then using that address to identify the likely owner
of the device), we did not do that in any of our experimentation.

Nonetheless, given the potential sensitivities of user locations our techniques
are able to infer, we took the following additional steps to respect users and
protect their privacy.

We ensured that all of our data was stored on machines that only the
researchers had access to.
We will make our code and aggregate statistics publicly available, but no
individual data---neither the scraped auctions nor the inferred locations.
This data management strategy factored into our IRB's determination that the
information we collected would not be used to personally identify any of the
users whose devices we studied in our experiments.

We followed \ebay's security researcher guidelines and did not engage in
research that could directly cause damage to \ebay users, systems, or
applications~\cite{ebay-sec}. 
We limited the rates at which we crawled and
queried their website to minimize any potential harm to users of the site and to
\ebay's systems.
We believe that our research to identify and help mitigate potential security
and privacy risks adheres to the principle of beneficence outlined in the Menlo
report~\cite{menlo}, and outweighs the potential risk to \ebay users.

Further, we have initiated the disclosure process to eBay, but as of the time of
this writing have yet to receive a reply.
We considered informing the individual users whose locations we were able to
track, but ultimately decided against this to protect the researchers.
Moreover, because eBay redacts expired auction pages, we believe that
by the time this work is made public, the affected users' pages will no
longer be available. 
Finally, we propose several mitigations in
\S\ref{sec:recommendations} that we believe will improve the security and
privacy of future \ebay users.

\section{Dataset Statistics}
\label{sec:results}

\begin{table}[t]
    \caption{Top \wifi AP manufacturers derived from \ebay AP photographs
    containing \acp{BSSID}.}
    \label{tab:manufs}
\begin{tabular}{r|r|c}
\multicolumn{1}{c}{\textbf{\# MAC Addresses}} & \multicolumn{1}{c}{\textbf{\%}} & \textbf{Manufacturer} \\
\hline
31,820 & 20.0 & Netgear \\
17,768 & 11.1 & Cisco \\
11,451 & 7.2 & Belkin \\
11,132 & 7.0 & TP-Link \\
8,069 & 5.1 & Ubiquiti \\
79,133 & 49.7 & 2,536 other \\
\hline
159,373 & 100 & \textbf{Total}
\end{tabular}
\end{table}

Between September 2024 and January 2025, we retrieved 144k \ebay auctions with
photographs containing MAC addresses (as identified through our \ac{OCR} pipeline
in \S\ref{sec:methodology}) from a total population of 788k auctions.  Some
auctions contained more than one string identified as a MAC address by our
\ac{OCR} pipeline. This resulted in slightly more (159k) unique MAC
addresses than auctions, for an average of 1.1 MAC addresses per auction.

Most of the auctions we identified did not result in a sale. To detect whether
the auction had a winner or not, we scraped each auction page for the phrase ``This
listing sold,'' which appears on listings that resulted in sale. %
While \ebay maintains the auction page for a period of time after an
item is sold or delisted, the site eventually removes the auction listing, and
at that point, we can no longer determine definitively whether an item was sold.
Of the 144k unique auctions we identified, we verified that 8,879 (6.2\%)
resulted in a sale through the ``This listing sold'' banner. In
\S\ref{sec:sold}, we consider the subset of items we are certain were sold in
our analysis. 
 
Using the \ac{MAC} addresses extracted from our \ebay \wifi AP auction \ac{OCR}
pipeline, we turn next to the task of geolocating these devices using
publicly-available geolocation data. First, we present some global and regional
statistics about our extracted \ac{MAC} address corpus. Then, using these
extracted \ac{MAC} addresses, we attempt to geolocate the devices before and
after auctions to determine whether \ebay listings containing MAC addresses pose
a unique threat to user privacy.

\subsection{Photograph-Derived MAC Address Corpus}
\label{sec:corpus}

Of the 159k MAC addresses produced by our
\ac{OCR} pipeline, common home \wifi AP brands dominate. Using the IEEE's \ac{OUI}
database~\cite{oui} to resolve the first three bytes of the MAC addresses we extract from
the \ebay photos, the most common router manufacturers are Netgear, Belkin,
Cisco, Ubiquiti and TP-Link.  Table~\ref{tab:manufs} lists the most
commonly-observed manufacturers in our corpus. %

\begin{table}[t]
    \caption{The number of \ebay auctions per seller country in our corpus.}
    \label{tab:countries}
\begin{tabular}{r|r|c}
\multicolumn{1}{c}{\textbf{\# Auctions}} & \multicolumn{1}{c}{\textbf{\%}} & \textbf{Country} \\
\hline
105,489 & 73.1 & US \\
20,228 & 14.0 & GB \\
4,684 & 3.2 & DE \\
3,426 & 2.4 & CN \\
2,666 & 1.9 & AU \\
7,864 & 5.5 & 142 other countries \\
\hline
144,357 & 100 & \textbf{Total}
\end{tabular}
\end{table}

The \ebay API lists the city, country, and partial postal code of the seller so
that prospective buyers can estimate shipping to their location. The majority of
the auctions that we enumerate through the \ebay API are listed by sellers in
the US. Table~\ref{tab:countries} lists the seller countries of the 144k
auctions we identified.

\begin{figure*}[t]
  \centering
    \begin{subfigure}[t]{0.45\linewidth} 
      \includegraphics[width=\linewidth]{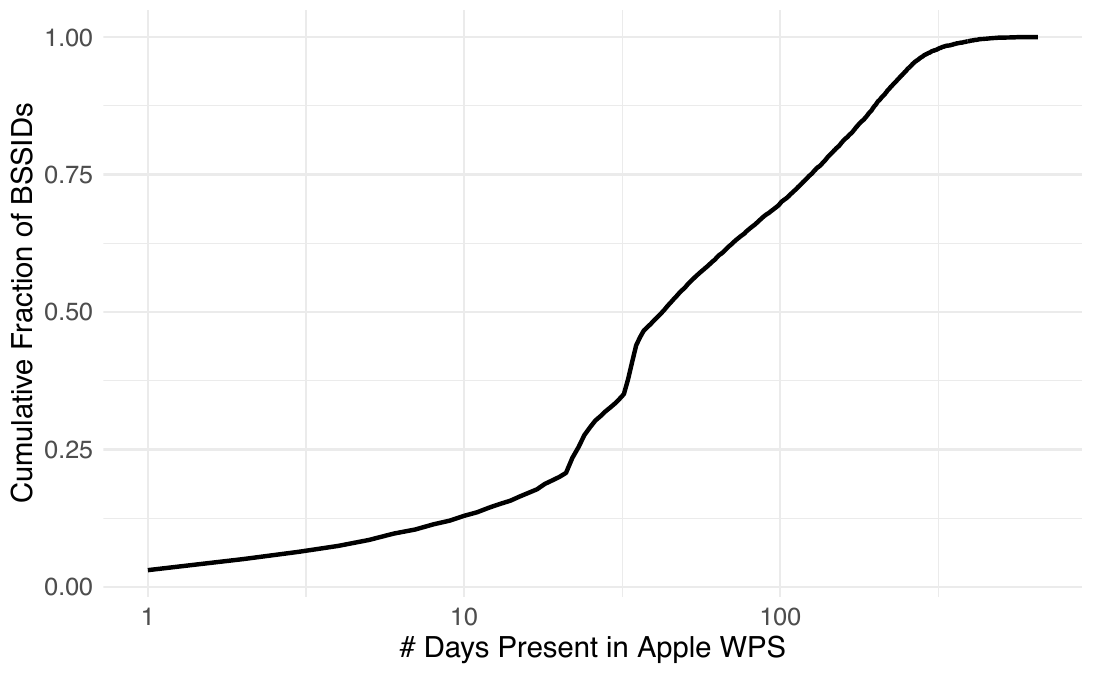} 
        \vspace{-5mm}
        \caption{Number of days observed in Apple's WPS as a cumulative fraction
        of BSSIDs ($x$-axis logscale).}
      \label{fig:wps-days}
    \end{subfigure}
    \hfill
    \begin{subfigure}[t]{0.45\linewidth} 
    \includegraphics[width=\textwidth]{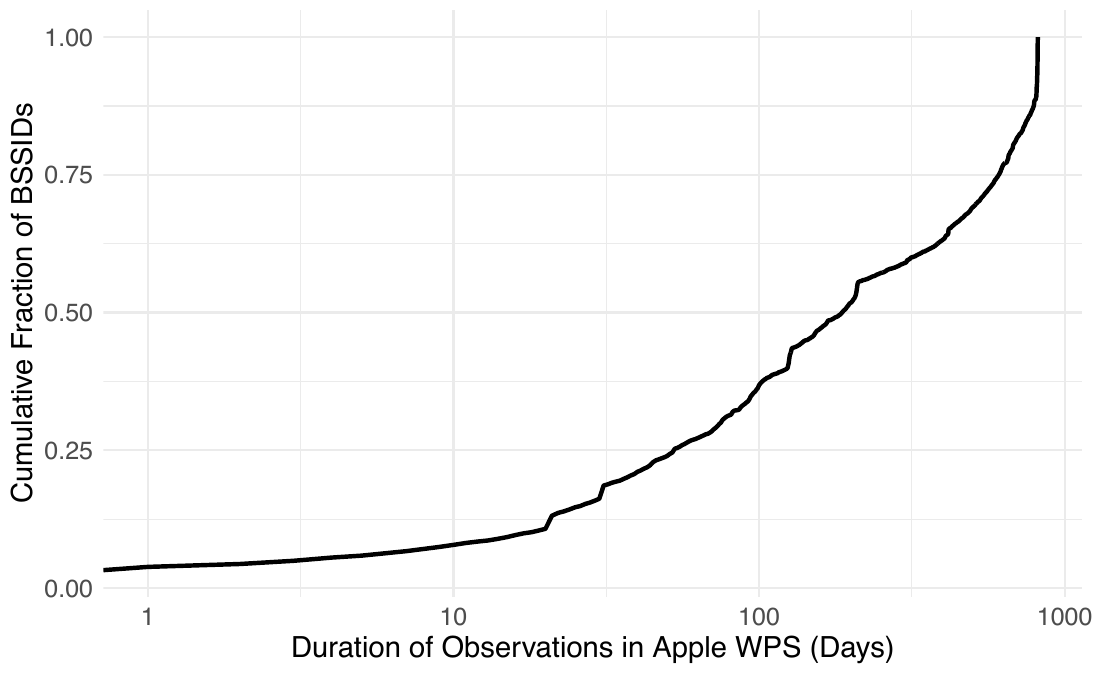} 
    \vspace{-5mm}
        \caption{Time between the first and last observation in the
        WPS as a cumulative fraction of BSSIDs ($x$-axis logscale).}
      \label{fig:wps-duration}
    \end{subfigure}
    \caption{Statistics for BSSID geolocations in Apple's WPS from all \ebay
    auctions.}
    \vspace{-5mm}
  \label{fig:bssid-stats}
\end{figure*}

\subsection{Extracted BSSID Geolocations}
\label{sec:results:geo}

To geolocate MAC addresses we extracted from \ebay auction photos, we relied on Apple's
\ac{WPS}.  Apple's \ac{WPS} is closed-source and the precise mechanics of the
system are not publicly disclosed; however, Rye and Levin discovered over 2B
\acp{BSSID} on all seven continents through the \ac{WPS} in their
study~\cite{rye2024surveilling}. Apple provides an accuracy value for geolocated
BSSIDs that typically ranges between 20-30m, and in testing, we experimentally
determined that it takes approximately 30 days for a previously-offline \ac{AP}
to become a landmark.

For this work, we used Apple's \ac{WPS} to look up the 159k MAC addresses we extracted from
\ebay auction photographs.  In the event that Apple's \ac{WPS} does not have a
record for a MAC address that we query (\eg, because it is not a BSSID on an
\ac{AP} or because the \ac{AP} has been turned off for some time), it returns an
invalid latitude and longitude (-180,-180). We continued to look these MAC
addresses up every day for a month, from mid-January to mid-February 2025.

In addition, we also looked these MAC addresses up in a longitudinal database of
BSSID geolocations dating to late 2022. This longitudinal data provides a
historical snapshot of where items from our \ebay corpus may have previously been. 
However, because this database was constructed without knowing the
\ebay MAC addresses \emph{a priori}, there is no guarantee that \ebay MAC
addresses are present in the historical data even if they were in-use BSSIDs.
Nonetheless, they may have been incidentally obtained and thus provide useful
historical location information.

Of the 159k distinct MAC addresses we extracted from photos, 13k (8.3\%) had
been geolocated at some point by querying the Apple \ac{WPS}.
Figure~\ref{fig:bssid-stats} highlights some overall features of the subset of
extracted MAC addresses that have BSSID geolocations in Apple's \ac{WPS}. In
particular, Figure~\ref{fig:wps-days} depicts the number of days BSSIDs were
geolocated in Apple's \ac{WPS}. The median number of days we observed BSSIDs in
the WPS was 43; the top 10\% of BSSIDs were observed in Apple's \ac{WPS} on 217
or more unique days. Figure~\ref{fig:wps-duration} displays cumulative fraction
of BSSIDs plotted against the duration of time they were observed in Apple's
WPS. That is, for each BSSID, we computed the length of time (in days) between
the first and last observations in Apple's WPS. The median timespan geolocated
BSSIDs were observed in the WPS was about six months at 187 days. The top 10\%
of BSSIDs had an observation window of more than 2 years, at 810 days.

Of the 13k BSSIDs we geolocated using Apple's WPS, 1,360 (10\%) were
geolocated \emph{only after} we began querying the MAC addresses extracted from
our \ebay photo corpus; they were not present in our longitudinal WPS
dataset prior to our extraction of the MAC address from an \ebay photograph.
This demonstrates that using \ebay as a source of \wifi MAC addresses yields a
significant number of new MAC addresses, despite having a 2-year longitudinal
corpus of BSSID geolocation data.

\section{Auction Geolocation Results}
\label{sec:auction-geo}

A major result of this work is the demonstration that an adversary may, with the
aid of machine learning techniques and Apple's \ac{WPS}, automate determining
the precise locations of the buyer and seller in an online auction involving
commodity \wifi \acp{AP}. In this section, we describe the situations in which
we are able to identify the parties involved in \ebay auctions, and how we
validate those geolocations where possible.

Of the 13k WPS-geolocated BSSIDs from our \ebay auction photograph corpus, there
are three distinct categories these devices may fall into, depending on when
they were geolocated relative to the time of their auction. We consider each of
these categories in turn.

\subsection{Geolocating \ebay Sellers}

The largest category of geolocated \ebay auction BSSIDs is comprised of BSSIDs
that were geolocated \emph{only before} the \ebay auction was listed.  These
5,628 BSSIDs may have been geolocated at any time prior to their listing; our
observation window of BSSIDs stretches back to late 2022.

Because these BSSIDs were geolocated \emph{only before} the auction occurred,
the geolocation obtained by Apple's \ac{WPS} is \emph{potentially} the location
of the item seller. In some cases, however, an item may have changed owners
before being sold at auction, or the owner may have moved locations between last
having the \wifi AP online (and thus discoverable via the Apple WPS) and selling
it.

To validate that we are, in many cases, discovering the precise location of
auction seller, due to their having used the \wifi AP at their home or business,
we perform a limited validation using the coarse-grained geolocation information
\ebay provides through its listing. For the US, this consists of the first three
digits of the five digit postal code.

Of the 4,311 geolocated \ebay BSSIDs whose auction listed the seller as in the
US, we compared the full US postal code derived from the latitude and longitude of
the Apple WPS location with the three-digit postal code prefix provided by
\ebay. Of these, only 4,085 had US postal codes and were thus able to be
compared. The rest were geolocated outside the US. We defer a closer examination
of some of these items to \S\ref{sec:casestudies}.

In approximately 50\% of cases (2,031) the postal code derived from the Apple
WPS geolocation matched the three-digit postal code prefix provided by \ebay.
Further, two-thirds (2,766, 67\%) of the postal codes derived from Apple's WPS
geolocations matched the first two digits of the \ebay-provided postal code
prefix, which serves as a coarser-grain match, and may account for moves the
seller made since using the item, or selling the item through a service or
dealer. 

Our technique is not limited to US \ebay sellers. Of the auctions in which the
BSSID was only geolocated before the listing, 895 BSSIDs were found in
photographs from auctions with a UK seller. Of these, 710 had geolocation
coordinates from Apple's WPS that indicated they were in the UK; from these
coordinates, we derived the UK postal code. When we compared the postal code
from our Apple WPS geolocations, 341 (48\%) of the WPS-derived postal codes matched the
postal code provided by the \ebay API.

This demonstrates that in many cases \ebay photographs can, via Apple's WPS,
leak fine-grained geolocation information about the seller of the item. 

\subsection{Geolocating \ebay Buyers}

In our corpus of photograph-derived BSSIDs, 3,123 were geolocated in Apple's WPS
\emph{only after} the \ebay auction was listed. In these instances, we are
likely geolocating the buyer of an item if it was sold on \ebay, or the new
owner of the item if it was delisted from \ebay and sold or donated through
another medium. 

Unlike with sellers, \ebay provides no location metadata about the \emph{buyer}
of an item, and it is unaware of the locations of new owners who may have
obtained the item outside of its platform. As such, we have nothing to compare
our post-auction geolocations to, other than to the seller's location. 

Intuitively, items sold on \ebay are relatively unlikely to be sold to a buyer
in the same postal code as the seller, and the fraction of the Apple WPS
geolocations of \ebay-derived postal codes for items geolocated \emph{after} the
auction should be much lower than the geolocations of items \emph{before} the
auction was listed. 

In contrast to the items geolocated in Apple's WPS prior to their listing on
\ebay, BSSIDs on items geolocated \emph{after} their listing on \ebay were
rarely in the same postal code as the seller. When we considered the three-digit
postal code prefix \ebay reports for US sellers, only 139 of 2,222 (6.3\%) BSSIDs
geolocated to the US via Apple's WPS matched the \ebay-listed postal code prefix.  This
rate of WPS-derived BSSID postal code matching is approximately eight times less
than the rate at which postal codes matched devices geolocated \emph{before} the
auction started.

This phenomenon is not limited to the US, either. Among auctions with a BSSID
geolocated to the UK (403 total), only 17 (4.2\%) BSSIDs geolocated to the same
postal code listed by \ebay. This validates our assumption that when devices are
geolocatable only after they appear in an \ebay auction, they have likely been
sold (whether on \ebay or not), and are frequently sold outside of the seller's
postal code.

\begin{figure}[t]
  \includegraphics[width=\linewidth]{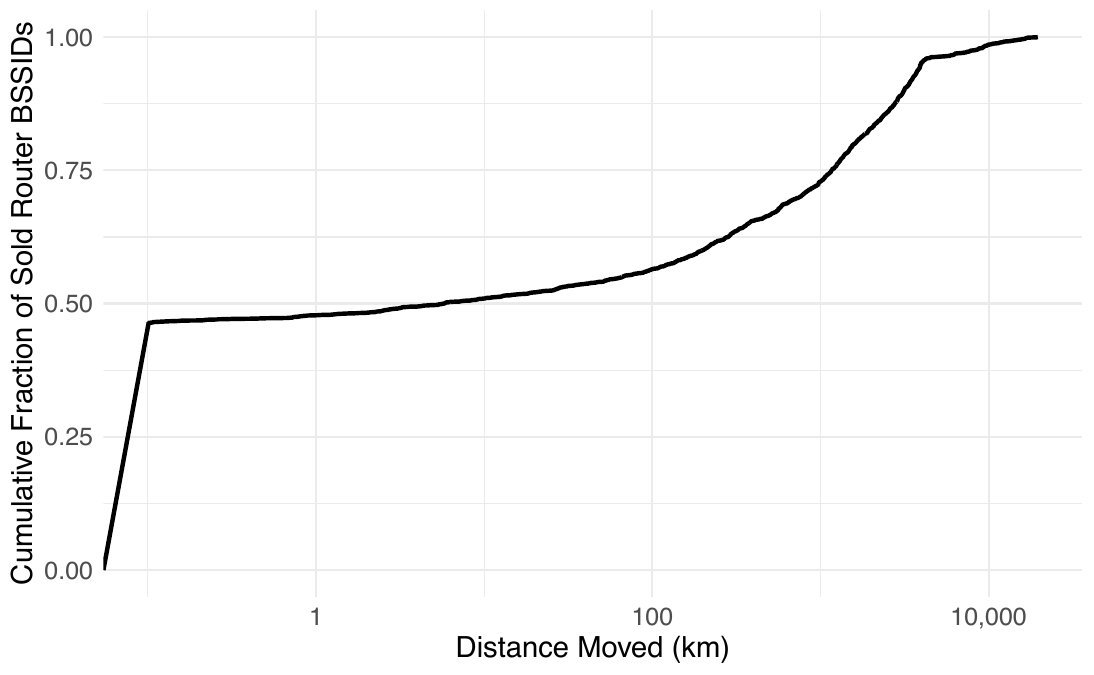}
    \vspace{-5mm}
    \caption{Distance moved by \wifi AP BSSIDs from auctions in which the AP 
    sold ($x$-axis logscale).}
  \label{fig:both-geo-moved}
\end{figure}

\subsection{Geolocating \ebay Buyers \emph{and} Sellers}
\label{sec:sold}

The worst outcome from a privacy perspective is that \ebay photographs for an
auction listing leak both the seller's location, from past geolocations,
\emph{as well as} the buyer's location after it is purchased and installed at
their location. We attempt quantify this phenomenon by identifying the \ebay
auctions we learned about that resulted in the sale of the device.

Of the 8,879 auctions that we determined were sold (\S\ref{sec:results}) 458
(5.2\%) auctions contained 439 distinct BSSIDs that were
geolocatable in Apple's WPS both before and after the auction occurred. In order
to validate that these devices were indeed sold (and thus likely moved), we
calculate the distance between the first and last geolocations for each of these
BSSIDs.

Figure~\ref{fig:both-geo-moved} plots the distances moved by cumulative fraction
of sold AP BSSIDs. The plot is bimodal; there is a first mode between 0 and
1 kilometers moved, indicating that the AP, which was ostensibly sold on
\ebay, did not actually move locations at all. While initially surprising, we
discovered two reasons for this phenomenon. First, some auctions list photos of
a device that is not actually the one sold. For instance, many bulk sales of
\wifi APs may list a close-up photograph of a single exemplar device, while many
individual devices may be sold under the same listing. Second, some 
vendors appear to reuse MAC addresses. Thus, while one device with a specific
BSSID may be online and appear stationary in Apple's WPS before and after the
auction, there is another device with an identical BSSID being sold during the
same time period.

Roughly half of the BSSIDs we extract from \ebay photos do not move
significantly. 
The remainder move nontrivial distances
between their initial and final observations in our WPS corpus. This
demonstrates that in hundreds of cases, we are able to geolocate both the seller
and buyer of \ebay APs via BSSIDs extracted from the auction photos.

\section{Case Studies}
\label{sec:casestudies}

In this section we detail several case studies that highlight the additional
information that can be gleaned from \ebay auctions that include visible MAC
addresses in their photos.

\subsection{Devices from Conflict Areas} %

The \ebay listings we gathered included some devices that had recently been in
conflict zones.  Our listings include two Kasa Smart Plugs, \wifi-enabled
electrical plugs that can be turned on and off through the companion smartphone
application~\cite{kasa}, that were present in Gaza at the time of the October
7th, 2023 Hamas attacks and subsequent Israel-Hamas War.  While the smart plugs
were in Gaza, they were approximately 10 kilometers apart.

On \ebay, these devices were listed by two different accounts on opposite sides
of the US.  How these devices came to be listed by these two sellers is unknown,
as both of the devices are stationary within Gaza and have no other geolocations
in the WPS.

\subsection{Auctioned Devices in Sensitive Locations} %

Some of the \ebay auction photographs we collected contained BSSIDs that were
later geolocated at particularly sensitive locations, such as military
installations. Using US military base location data~\cite{millocations}, we
determined whether any of the geolocated BSSIDs derived from \ebay photos were
within their boundaries.

Thirteen of the BSSIDs from \ebay photos that we geolocated with Apple's WPS
were on US military bases. Of these, 5 were geolocated \emph{after} the \ebay
auction occurred, indicating that they were likely purchased by residents or
organizations located on the military base. 

This case study presents a potential attack vector for attackers---sell
vulnerable \wifi access points in online forums, such as \ebay, and wait for
them to appear as landmarks in a WPS. If they are at a potentially sensitive
location, such as a military base, university, or hospital, the attacker then
knows about a vulnerable network device that can be used to gain a foothold in
the buyer's network.

\begin{figure*}[t]
  \centering
    \begin{subfigure}[t]{0.48\linewidth} 
        \includegraphics[width=\linewidth]{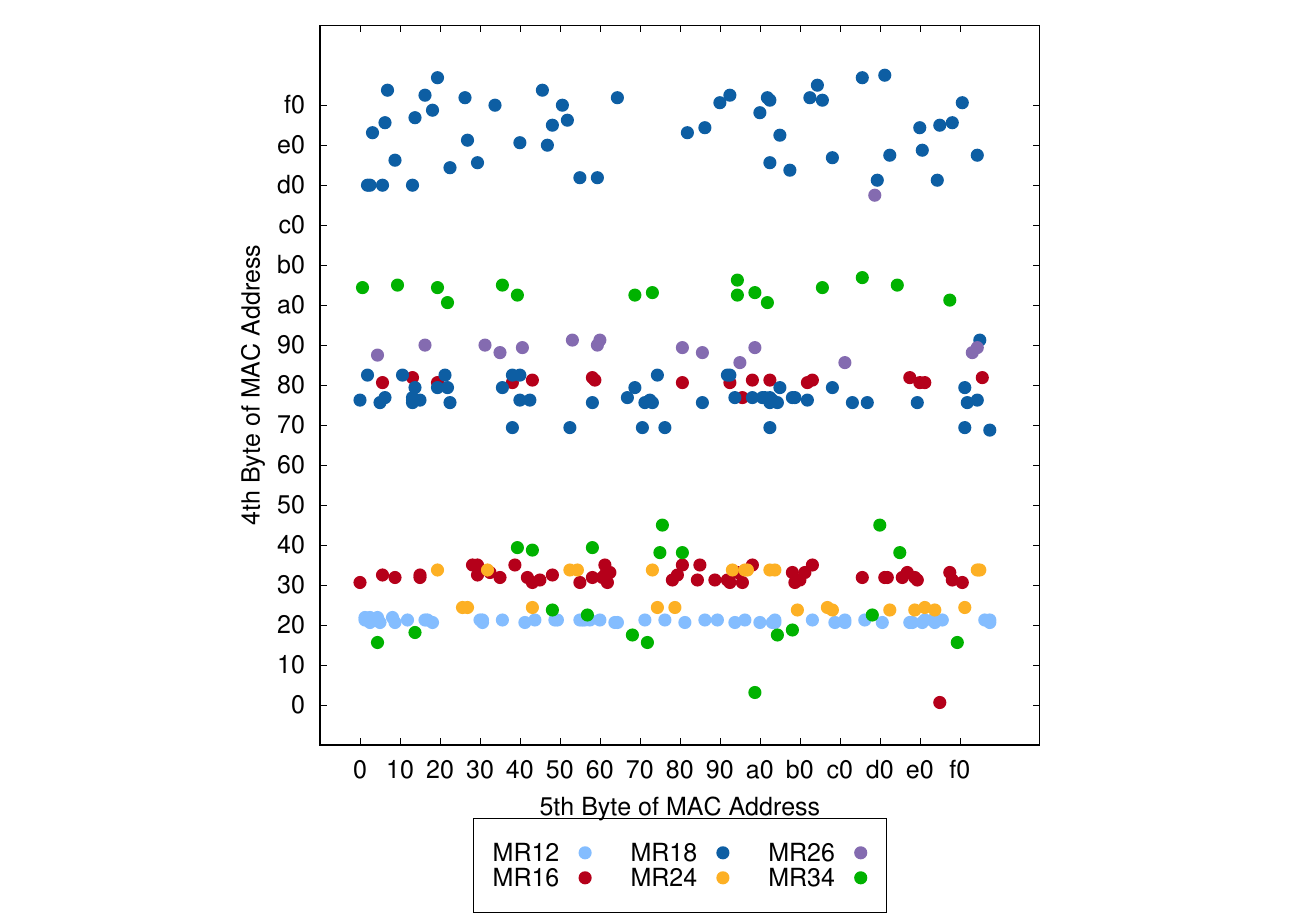}
        \caption{Model observations from a Cisco Meraki OUI
        (\texttt{00:18:0a}).}
      \label{fig:meraki-scatter}
    \end{subfigure}
    \hfill
    \begin{subfigure}[t]{0.48\linewidth} 
        \includegraphics[width=\textwidth]{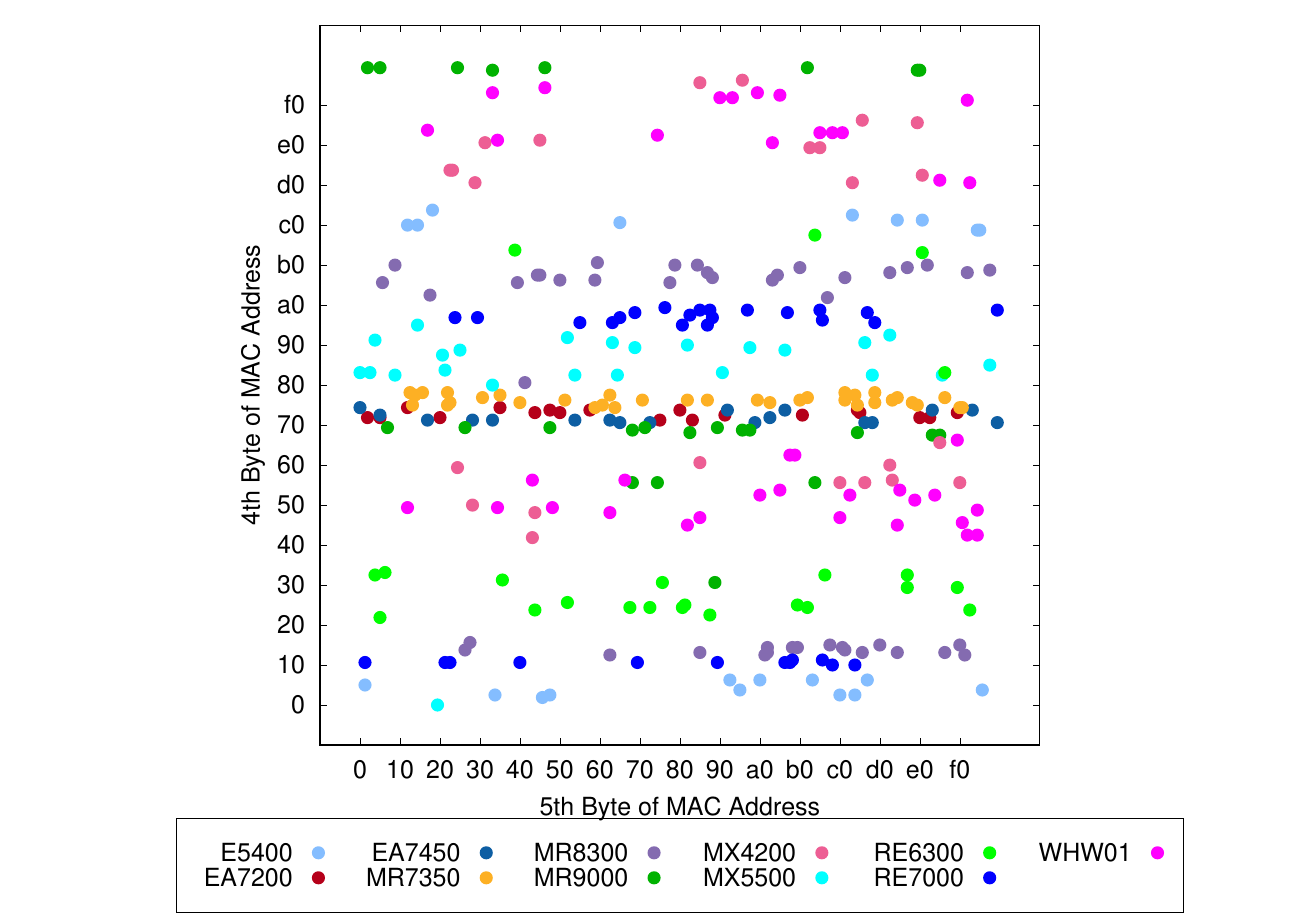}
        \caption{Model observations from a Belkin OUI (\texttt{e8:9f:80}).}
      \label{fig:belkin-scatter}
    \end{subfigure}
    \caption{Plots of device models for two different manufacturer's OUIs.
    Different color points indicate different model types.}
    \label{fig:model-inference}
\end{figure*}

\subsection{Used Items Sold as New} %

We investigated whether \ebay sellers ever miscategorize the condition
of their auctioned goods by indicating that devices are new when they have in fact been
used previously. \ebay has different categories for different types of items
(\eg, clothes have a ``new with tags'' category, while computer equipment does
not); we consider both ``new'' and ``open box''---which \ebay describes as
``excellent, new condition with no wear''~\cite{conditions}---labels and check
our longitudinal WPS data for evidence of the device operating before the
auction took place.

Our data shows that previously-used \wifi APs are regularly mischaracterized as
``new''. Of the 13k \ebay photo-derived BSSIDs, 1,089 (8.1\%) were observed in
our Apple WPS corpus \emph{before} the auction was listed, and the auction
described the \wifi AP's device type as ``new'' or ``open box''. The majority
(684, 63\% of the mischaracterized BSSIDs) of these mislabeled auctions listed
their device as ``open box'', rather than an unqualified ``new''. We speculate
that the sellers may have chosen ``open box'', rather than ``new'', to provide
a plausible explanation for any wear or damage to the devices incurred while it
was actually in operation. 

As an example, in late 2024 a seller from the US listed a lot of eight Aruba
enterprise wireless APs with the condition ``open box.'' However, a MAC address
displayed in one of the photographs (a MAC address on only one of the eight
routers was visible) was geolocated to a university campus in the Middle East
via Apple's WPS for most of 2023. This seller's auction feedback, which is
publicly visible, lists several negative comments from \ebay users complaining
that the seller auctions used electronics equipment that they have salvaged.

\subsection{Granular Model Inferences} %
\label{sec:model-inference}

In our final case study, we demonstrate another attack that our corpus has made
possible: inferring the specific model of devices from knowing only their MAC
address.
Such \emph{model inference attacks} can be valuable to attackers in scenarios
where the attacker is able to learn a target's MAC address, and wishes to use a
model-specific exploit.
As discussed in \S\ref{sec:background}, prior work on model inference required
communicating directly with devices using obsolescent protocols.
Here, we show that Martin~\etal's results~\cite{martin2016decomposition} can be
replicated remotely by extracting MAC addresses using OCR.

We emphasize that, for devices listed on \ebay, this attack would have little utility,
as the specific model could be inferred from the description or photo.
Rather, this attack is for scenarios where the attacker learns a target's MAC
address through some other means (e.g., inclusion in an IPv6 EUI-64 address).

In addition to any MAC addresses extracted as part of our OCR pipeline
described in \S\ref{sec:methodology}, we also extract manufacturer and model
information, when present, from text on the listing title. When both pieces of
information are present, we create a linkage between the MAC address,
manufacturer, and model listed on the device.

Figure~\ref{fig:model-inference} depicts the model observations across two OUIs
from two different manufacturers, and highlights vendor MAC address allocation
strategy differences between them. Figure~\ref{fig:meraki-scatter} plots the
fourth and fifth bytes of the MAC addresses of devices with the
\texttt{00:18:0a} OUI; each color represents a different model as indicated by
the legend. Most models have distinct banding patterns, as evidenced by points
of the same color being grouped in layers across fourth-byte values.
Figure~\ref{fig:belkin-scatter} also exhibits banding, albeit with more than
twice as many models observed within the OUI, and with tighter bands.

\section{Limitations}
\label{sec:limitations}

While the preceding results demonstrate the real-world feasibility
and broad applicability of our attack, there are several current and
future potential limitations that bear discussion.

First, 
a potential source of error lies in the fallibility of the \ac{OCR} \ac{MAC}
address extraction pipeline.  When a MAC address is present in the 
user-supplied photograph, the pipeline may fail to detect
its presence, or may infer an incorrect address, for one of several
reasons.
First, the image may be blurry, contain
the text of the \ac{MAC} address at an
indirect angle causing occlusion or distortions,
or have a resolution insufficient to
distinguish individual characters. For instance, the valid
hexadecimal characters ``\texttt{B}''
and ``\texttt{8}''
are often difficult---even for a human---to differentiate in low
resolution photos or when the image is taken at an angle. 
In other cases, the MAC address may be
truncated due to the framing of the photo or otherwise obscured. Last, it can sometimes be difficult 
to differentiate MAC addresses from other numeric identifiers such as
serial numbers due to the variety of representation formats
manufacturers employ,
\eg, use of a colon, dash, space, or no separator
at all to mark digits, octets or pairs.  
To quantify the extent to which such errors affect our pipeline in 
practice, we validate our pipeline's MAC address extraction results against human annotation
in~\S\ref{sec:macvalidation}.

A second potential confounding factor
involves cases in which the equipment sold is
not the same as the equipment depicted in the auction image. For
instance, 
some \ebay
auctions are lots---an auction in which many of the same item is for
sale. The image depicted in a lot listing may thus
contain a valid \ac{MAC} address, but for a single unit of the multiple-item lot.
Similarly, it is possible that for an auction of a single unit, the seller may
include a picture of a device different from the one that they ship to the
eventual buyer, for instance by using a ``stock'' image.

Naturally, our attack depends on the presence of MAC addresses in the
auction listing image.  
We are unaware of a legal or regulatory requirement for addresses to be
printed on a wireless access point or its packaging, and presumably some manufacturers omit this
information.  In this case, our \ac{OCR} \ac{MAC} address extraction pipeline
will fail to produce a valid \ac{MAC} address because none exists.
Alternatively, it is possible that the seller of the device may neglect to take
a picture of that part of the device in the event that a \ac{MAC} address
\emph{is} listed. For some rare cases, we have observed sellers redacting MAC
addresses in their listing photos along with other sensitive information.
In either of these cases, it is impossible for our pipeline to
extract the correct MAC address for an auction---indeed,
we advocate for automatically redacting addresses as a potential 
mitigation mechanism in~\S\ref{sec:recommendations}.

A fourth source of potential error stems broadly from correlation
with the \ac{WPS} data.  A MAC address may not be present in Apple's
WPS database if it is unstable, moving, or short-lived (and, hence,
does not serve as a valuable landmark for their service).  Further,
Apple's data is dynamic and even if the MAC address were present at 
one point in time, it may not be within the data we query or the available 
historical data.  We utilize historic data and quantify the extent
to which we successfully find addresses in Apple's database
in~\S\ref{sec:results:geo}.  Of course, the correlation attack depends trivially on
the availability of the WPS database itself.  Should Apple elect to restrict
access to, or protect the privacy of, the content in the database---as 
recommended by~\cite{rye2024surveilling}---our location tracking will cease to be
effective.  

Related to the prior limitation, the attack depends on persistent MAC
addresses.  While most client devices employ MAC
randomization~\cite{fenske2021three}, this behavior has not been widely adopted in
\wifi \acp{AP}.  An \ac{AP} could choose random short-lived BSSIDs when
sending beacon frames advertising their SSID, yet, in practice, few
do.  However, as awareness
of the sensitivity of MAC addresses with respect to privacy and
tracking increases, we may reasonably expect vendors to adopt
randomization.

\subsection{Generalizability}
\label{sec:generalizability}

This study focused on the \ebay marketplace due to its available and
well-documented API. To understand how generalizable our attack is to other
platforms, we performed a small-scale experiment 
to examine the efficacy of our MAC address extraction pipeline 
on Craigslist and Facebook
Marketplace.

Both Craigslist and Facebook Marketplace localize their results to the user's
location with the intention of facilitating in-person sales and
transactions.
For consistency, we localized results near our academic institution on 
both sites. We used the search term ``wifi router''---one of the general search
queries we used in our \ebay data collection---and downloaded all
available images for approximately the top 100 listings from
each website. For Facebook Marketplace, which permits sellers to upload a video,
we retrieved the video thumbnail. %

On Craigslist, our OCR-based extraction identified 17 of 113 listings (15\%)
as containing images with \ac{MAC} addresses. %
Note that Craigslist permits listings without photos; 13 of the 113 listings
(11.5\%) had no associated images.
On Facebook Marketplace, our OCR extraction identified 29 of 135 listings (21\%)
with images containing \ac{MAC} addresses.
Manual inspection of the extracted MAC addresses from both platforms indicated 
an accuracy 
comparable to our more in-depth validation
performed with the \ebay images in \S\ref{sec:macvalidation}.

These MAC address extraction percentages (15\% and 21\%) are roughly in line
with our \ebay MAC address extraction rate, in which approximately 18\%
($\sim$$\frac{188,000}{788,000}$) of listings had images with
\ac{MAC} addresses present.  The lower hit rate of Craigslist is in
part attributable to the fact that listing images are optional.

Similar to \ebay auctions, we also observed some Craigslist and Facebook
Marketplace listings that contained multiple \ac{MAC} addresses. On Craigslist, 19
unique \ac{MAC} addresses were extracted from 17 listings; on Facebook
Marketplace, 43 MAC addresses were extracted from 29 listings.

The ethical considerations of this validation study largely mirror those
described in \S\ref{sec:ethics}. In addition, we collected our limited
validation photos from both sites by manually browsing to retrieve listing
images, rather than programatically scraping or spidering the sites. While
Craigslist's Terms of Use~\cite{cltou} prohibit copying of Craigslist data ``by
hand'', our small-scale validation study is indistinguishable from normal
browsing. %
Finally, we have initiated the responsible disclosure process
with Craigslist and Facebook to inform them of our findings.

In sum, while our study exclusively focuses on \ebay, our preliminary results
show that when users list \wifi \acp{AP} for sale, they frequently include photos
containing MAC addresses irrespective of platform. Further, our experimentation
with Craigslist and Facebook Marketplace demonstrate that that our OCR pipeline
is able to extract these MAC addresses, as well.

\section{Recommendations}
\label{sec:recommendations}

We propose several solutions that will mitigate the threat posed by posting
photos online of \wifi AP device identifiers.

\parhead{Policy Modifications} 
\ebay and other online forums can advise users \emph{not} to post photographs
of sensitive device information. At the time of this writing, \ebay encourages
sellers to post many, detailed photographs of their item for
sale~\cite{ebay-picture-tips}, without warning them that particular information
might compromise their location privacy. A warning explaining the privacy
threat posed by uploading images with sensitive device data may deter users
from doing so. 

Among our three recommendations, this is the easiest to deploy---it would only
involve eBay changing some text on their website---but likely the least
effective, because it relies completely on users taking the appropriate action.

\parhead{Image Manipulations}
If users do upload images containing device identifiers, \ebay should manipulate
those photos to obscure sensitive information, such as the BSSID or other MAC
addresses, \acp{SSID}, and passwords, that are commonly printed on the backs of
\wifi APs. Though used for different purposes, mapping websites like Google Maps
blur sensitive information like license plates and peoples' faces. \ebay and
other online forums should employ a similar method of redacting sensitive
information using an OCR pipeline similar to the one we employed; after
identifying areas that contain sensitive information, a blur or opaque box could
obscure the data below.

Compared to our other recommendations, image manipulations would be moderately
difficult to deploy, as it requires deploying additional code at a site like
eBay.
Additionally, the computational burden of processing the volume of images \ebay's
platform receives through a vision pipeline similar to the one we develop in this
paper may present a challenge. However, further refinements for efficiency such as 
pre-filtering of images with less intensive methods can alleviate this barrier.
It would be more effective than policy changes, but would be limited by the OCR
pipeline's accuracy---which, while effective in the typical conditions we
explored (clear images of a single AP) may be less effective in conditions we
did not test for (e.g., if the image contained multiple APs).
One of the benefits of image manipulations is that it would apply also to legacy
APs that cannot deploy their own defenses.

\parhead{\wifi AP Protections}
\wifi AP manufacturers can protect users from location privacy threats by
refraining from using the BSSID printed on the device. Rye and
Levin~\cite{rye2024surveilling} recommend the use of \emph{random BSSIDs} to
prevent longitudinal tracking via a WPS; using a random MAC address prevents
targeted localization of a particular device through a known BSSID extracted
from an \ebay photo. 

Among our three recommendations, AP-based protections are by far the most
effective, as they preclude the possibility of attributing any given MAC
address to any picture of hardware.
However, it is also by far the most difficult to deploy; it would likely
require changes to the wireless specifications to ensure that AP manufacturers
adopt MAC randomization.
Worse yet, it is unclear whether MAC randomization is even desirable amongst
the entire networking community; network administrators commonly use static MAC
addresses to assist them in managing and tracking the devices on their network.
Future work is necessary to find the balance between privacy preservation and
facilitated network management.

\section{Conclusion}
\label{sec:conclusion}

In this work, we demonstrated that users' reasonable expectations of location
privacy before and after an online auction can be violated merely through the
MAC addresses included in the auction's public photos.
Our primary contributions include a data analysis pipeline for identifying and
extracting MAC addresses from photos in an automated manner, and then
cross-referencing these MAC addresses against a WPS dataset known to expose
device locations.
We were able to extract over 144k MAC addresses, identify the locations
of 5,628 sellers, 3,123 buyers, and an additional 458 buyer-seller pairs.
Our validation experiments show the efficacy of our techniques, and our case
studies demonstrate some of the potential severity of the attack, such as the
presence of secondhand network devices being used on US military bases.

In sum, our work shows that a seemingly innocuous photo of a MAC address
displayed on the side of a networking device can have surprisingly large
privacy implications.
We provide several recommendations for users, auction services, and network
device manufacturers to take to mitigate this attack.

\begin{acks}
    We are grateful for the anonymous reviewers and shepherd for their valuable
    feedback and suggestions during the revision process. This work was
    supported in part by NSF grants CNS-1943420 and CNS-2323193.
\end{acks}

\bibliography{conferences,refs}


\begin{thebibliography}{39}


\ifx \showCODEN    \undefined \def \showCODEN     #1{\unskip}     \fi
\ifx \showDOI      \undefined \def \showDOI       #1{#1}\fi
\ifx \showISBNx    \undefined \def \showISBNx     #1{\unskip}     \fi
\ifx \showISBNxiii \undefined \def \showISBNxiii  #1{\unskip}     \fi
\ifx \showISSN     \undefined \def \showISSN      #1{\unskip}     \fi
\ifx \showLCCN     \undefined \def \showLCCN      #1{\unskip}     \fi
\ifx \shownote     \undefined \def \shownote      #1{#1}          \fi
\ifx \showarticletitle \undefined \def \showarticletitle #1{#1}   \fi
\ifx \showURL      \undefined \def \showURL       {\relax}        \fi
\providecommand\bibfield[2]{#2}
\providecommand\bibinfo[2]{#2}
\providecommand\natexlab[1]{#1}
\providecommand\showeprint[2][]{arXiv:#2}

\bibitem[clt(2025)]%
        {cltou}
 \bibinfo{year}{2025}\natexlab{}.
\newblock \bibinfo{title}{{Craigslist -- Terms of Use}}.
\newblock
\newblock
\newblock
\shownote{\url{https://www.craigslist.org/about/terms.of.use/en}}.


\bibitem[eba(2025a)]%
        {ebay-picture-tips}
 \bibinfo{year}{2025}\natexlab{a}.
\newblock \bibinfo{title}{{eBay -- Adding pictures to your listing}}.
\newblock
\newblock
\newblock
\shownote{\url{https://www.ebay.com/help/selling/listings/adding-pictures-listings?id=4148}}.


\bibitem[eba(2025b)]%
        {ebay-sec}
 \bibinfo{year}{2025}\natexlab{b}.
\newblock \bibinfo{title}{{eBay -- Security Researchers}}.
\newblock
\newblock
\newblock
\shownote{\url{https://pages.ebay.com/securitycenter/security_researchers.html}}.


\bibitem[eba(2025c)]%
        {ebay-advanced-search-documentation}
 \bibinfo{year}{2025}\natexlab{c}.
\newblock \bibinfo{title}{{eBay Advanced Search Documentation}}.
\newblock
\newblock
\newblock
\shownote{\url{https://web.archive.org/web/20241210060542/https://developer.ebay.com/api-docs/user-guides/static/finding-user-guide/finding-searching-by-keywords.html}}.


\bibitem[eba(2025d)]%
        {ebay-api-limit-documentation}
 \bibinfo{year}{2025}\natexlab{d}.
\newblock \bibinfo{title}{{eBay Call Limit Documentation}}.
\newblock
\newblock
\newblock
\shownote{\url{https://developer.ebay.com/develop/get-started/api-call-limits}}.


\bibitem[eba(2025e)]%
        {ebay-search-documentation}
 \bibinfo{year}{2025}\natexlab{e}.
\newblock \bibinfo{title}{{eBay search API Documentation}}.
\newblock
\newblock
\newblock
\shownote{\url{https://developer.ebay.com/api-docs/buy/browse/resources/item_summary/methods/search}}.


\bibitem[kas(2025)]%
        {kasa}
 \bibinfo{year}{2025}\natexlab{}.
\newblock \bibinfo{title}{{Kasa: Creating a true smart home experience all in
  one app}}.
\newblock
\newblock
\newblock
\shownote{\url{https://www.kasasmart.com}}.


\bibitem[pad(2025)]%
        {paddleOCR}
 \bibinfo{year}{2025}\natexlab{}.
\newblock \bibinfo{title}{{PaddleOCR}}.
\newblock
\newblock
\newblock
\shownote{\url{https://github.com/PaddlePaddle/PaddleOCR}}.


\bibitem[qwe(2025)]%
        {qwenvl}
 \bibinfo{year}{2025}\natexlab{}.
\newblock \bibinfo{title}{{Qwen-VL Github}}.
\newblock
\newblock
\newblock
\shownote{\url{https://github.com/QwenLM/Qwen-VL}}.


\bibitem[Abedi et~al\mbox{.}(2013)]%
        {abedi2013bluetooth}
\bibfield{author}{\bibinfo{person}{Naeim Abedi}, \bibinfo{person}{Ashish
  Bhaskar}, {and} \bibinfo{person}{Edward Chung}.}
  \bibinfo{year}{2013}\natexlab{}.
\newblock \showarticletitle{{Bluetooth and Wi-Fi MAC Address Based Crowd
  Collection and Monitoring: Benefits, Challenges and Enhancement}}. In
  \bibinfo{booktitle}{\emph{{Australasian Transport Research Forum}}}.
\newblock


\bibitem[Bailey et~al\mbox{.}(2012)]%
        {menlo}
\bibfield{author}{\bibinfo{person}{Michael Bailey}, \bibinfo{person}{David
  Dittrich}, \bibinfo{person}{Erin Kenneally}, {and} \bibinfo{person}{Doug
  Maughan}.} \bibinfo{year}{2012}\natexlab{}.
\newblock \showarticletitle{{The Menlo Report}}.
\newblock \bibinfo{journal}{\emph{IEEE Symposium on Security and Privacy}}
  (\bibinfo{year}{2012}).
\newblock


\bibitem[Becker et~al\mbox{.}(2019)]%
        {becker2019tracking}
\bibfield{author}{\bibinfo{person}{Johannes~K Becker}, \bibinfo{person}{David
  Li}, {and} \bibinfo{person}{David Starobinski}.}
  \bibinfo{year}{2019}\natexlab{}.
\newblock \showarticletitle{{Tracking Anonymized Bluetooth Devices}}.
\newblock \bibinfo{journal}{\emph{Privacy Enhancing Technologies Symposium
  (PETS)}} (\bibinfo{year}{2019}).
\newblock


\bibitem[Bongard(2014)]%
        {bongard2014offline}
\bibfield{author}{\bibinfo{person}{Dominique Bongard}.}
  \bibinfo{year}{2014}\natexlab{}.
\newblock \showarticletitle{{Offline Bruteforce Attack on WiFi Protected
  Setup}}.
\newblock \bibinfo{journal}{\emph{Passwordscon}} (\bibinfo{year}{2014}).
\newblock


\bibitem[Celosia and Cunche(2020)]%
        {celosia2020discontinued}
\bibfield{author}{\bibinfo{person}{Guillaume Celosia} {and}
  \bibinfo{person}{Mathieu Cunche}.} \bibinfo{year}{2020}\natexlab{}.
\newblock \showarticletitle{{Discontinued Privacy: Personal Data Leaks in Apple
  Bluetooth-Low-Energy Continuity Protocols}}.
\newblock \bibinfo{journal}{\emph{Privacy Enhancing Technologies Symposium
  (PETS)}} (\bibinfo{year}{2020}).
\newblock


\bibitem[Costantin et~al\mbox{.}(2017)]%
        {costantin2017vulnerabilities}
\bibfield{author}{\bibinfo{person}{Daniel Costantin}, \bibinfo{person}{Krishnun
  Sansurooah}, {and} \bibinfo{person}{Patricia~AH Williams}.}
  \bibinfo{year}{2017}\natexlab{}.
\newblock \showarticletitle{{Vulnerabilities Associated with Wi-fi Protected
  Setup in a Medical Environment}}. In \bibinfo{booktitle}{\emph{Australasian
  Computer Science Week Multiconference}}.
\newblock


\bibitem[Cunche(2014)]%
        {cunche2014know}
\bibfield{author}{\bibinfo{person}{Mathieu Cunche}.}
  \bibinfo{year}{2014}\natexlab{}.
\newblock \showarticletitle{{I Know Your MAC Address: Targeted Tracking of
  Individuals using Wi-Fi}}.
\newblock \bibinfo{journal}{\emph{{Computer Virology and Hacking Techniques}}}
  (\bibinfo{year}{2014}).
\newblock


\bibitem[eBay(2025)]%
        {conditions}
\bibfield{author}{\bibinfo{person}{eBay}.} \bibinfo{year}{2025}\natexlab{}.
\newblock \showarticletitle{{Item Condition by Category}}.
\newblock
\newblock
\shownote{\url{https://www.ebay.com/help/selling/listings/creating-managing-listings/item-conditions-category}}.


\bibitem[Fenske et~al\mbox{.}(2021)]%
        {fenske2021three}
\bibfield{author}{\bibinfo{person}{Ellis Fenske}, \bibinfo{person}{Dane Brown},
  \bibinfo{person}{Jeremy Martin}, \bibinfo{person}{Travis Mayberry},
  \bibinfo{person}{Peter Ryan}, {and} \bibinfo{person}{Erik~C Rye}.}
  \bibinfo{year}{2021}\natexlab{}.
\newblock \showarticletitle{{Three Years Later: A Study of MAC Address
  Randomization In Mobile Devices And When It Succeeds}}.
\newblock \bibinfo{journal}{\emph{Privacy Enhancing Technologies Symposium
  (PETS)}}.
\newblock


\bibitem[Freudiger(2015)]%
        {freudiger2015talkative}
\bibfield{author}{\bibinfo{person}{Julien Freudiger}.}
  \bibinfo{year}{2015}\natexlab{}.
\newblock \showarticletitle{{How Talkative is Your Mobile Device? An
  Experimental Study of Wi-Fi Probe Requests}}. In
  \bibinfo{booktitle}{\emph{ACM Conference on Security and Privacy in Wireless
  and Mobile Networks (WiSec)}}.
\newblock


\bibitem[Garfinkel and Shelat(2003)]%
        {garfinkel2003remembrance}
\bibfield{author}{\bibinfo{person}{Simson~L Garfinkel} {and}
  \bibinfo{person}{Abhi Shelat}.} \bibinfo{year}{2003}\natexlab{}.
\newblock \showarticletitle{{Remembrance of Data Passed: A Study of Disk
  Sanitization Practices}}.
\newblock \bibinfo{journal}{\emph{IEEE Symposium on Security and Privacy}}
  (\bibinfo{year}{2003}).
\newblock


\bibitem[Goodin(2013)]%
        {goodin2013no}
\bibfield{author}{\bibinfo{person}{Dan Goodin}.}
  \bibinfo{year}{2013}\natexlab{}.
\newblock \showarticletitle{{No, this isn't a scene from Minority Report. This
  trash can is stalking you}}.
\newblock \bibinfo{journal}{\emph{Ars Technica}} (\bibinfo{year}{2013}).
\newblock


\bibitem[IEEE(2025)]%
        {oui}
\bibfield{author}{\bibinfo{person}{IEEE}.} \bibinfo{year}{2025}\natexlab{}.
\newblock \bibinfo{title}{{MAC Address Block Large (MA-L)}}.
\newblock
\newblock
\newblock
\shownote{\url{https://standards-oui.ieee.org/oui/oui.txt}}.


\bibitem[Kirillov et~al\mbox{.}(2023)]%
        {kirillov2023segany}
\bibfield{author}{\bibinfo{person}{Alexander Kirillov}, \bibinfo{person}{Eric
  Mintun}, \bibinfo{person}{Nikhila Ravi}, \bibinfo{person}{Hanzi Mao},
  \bibinfo{person}{Chloe Rolland}, \bibinfo{person}{Laura Gustafson},
  \bibinfo{person}{Tete Xiao}, \bibinfo{person}{Spencer Whitehead},
  \bibinfo{person}{Alexander~C Berg}, \bibinfo{person}{Wan-Yen Lo},
  {et~al\mbox{.}}} \bibinfo{year}{2023}\natexlab{}.
\newblock \showarticletitle{{Segment Anything}}. In
  \bibinfo{booktitle}{\emph{{IEEE Conference on Computer Vision}}}.
\newblock


\bibitem[Martin et~al\mbox{.}(2019)]%
        {martin2019handoff}
\bibfield{author}{\bibinfo{person}{Jeremy Martin}, \bibinfo{person}{Douglas
  Alpuche}, \bibinfo{person}{Kristina Bodeman}, \bibinfo{person}{Lamont Brown},
  \bibinfo{person}{Ellis Fenske}, \bibinfo{person}{Lucas Foppe},
  \bibinfo{person}{Travis Mayberry}, \bibinfo{person}{Erik~C Rye},
  \bibinfo{person}{Brandon Sipes}, {and} \bibinfo{person}{Sam Teplov}.}
  \bibinfo{year}{2019}\natexlab{}.
\newblock \showarticletitle{{Handoff All Your Privacy: A Review of Apple's
  Bluetooth Low Energy Continuity Protocol}}.
\newblock \bibinfo{journal}{\emph{Privacy Enhancing Technologies Symposium
  (PETS)}} (\bibinfo{year}{2019}).
\newblock


\bibitem[Martin et~al\mbox{.}(2017)]%
        {martin2017study}
\bibfield{author}{\bibinfo{person}{Jeremy Martin}, \bibinfo{person}{Travis
  Mayberry}, \bibinfo{person}{Collin Donahue}, \bibinfo{person}{Lucas Foppe},
  \bibinfo{person}{Lamont Brown}, \bibinfo{person}{Chadwick Riggins},
  \bibinfo{person}{Erik~C Rye}, {and} \bibinfo{person}{Dane Brown}.}
  \bibinfo{year}{2017}\natexlab{}.
\newblock \showarticletitle{{A Study of MAC Address Randomization in Mobile
  Devices and When it Fails}}.
\newblock \bibinfo{journal}{\emph{Privacy Enhancing Technologies Symposium
  (PETS)}}.
\newblock


\bibitem[Martin et~al\mbox{.}(2016)]%
        {martin2016decomposition}
\bibfield{author}{\bibinfo{person}{Jeremy Martin}, \bibinfo{person}{Erik Rye},
  {and} \bibinfo{person}{Robert Beverly}.} \bibinfo{year}{2016}\natexlab{}.
\newblock \showarticletitle{{Decomposition of MAC Address Structure for
  Granular Device Inference}}. In \bibinfo{booktitle}{\emph{Annual Computer
  Security Applications Conference (ACSAC)}}.
\newblock


\bibitem[Matte et~al\mbox{.}(2016)]%
        {matte2016defeating}
\bibfield{author}{\bibinfo{person}{C{\'e}lestin Matte},
  \bibinfo{person}{Mathieu Cunche}, \bibinfo{person}{Franck Rousseau}, {and}
  \bibinfo{person}{Mathy Vanhoef}.} \bibinfo{year}{2016}\natexlab{}.
\newblock \showarticletitle{{Defeating MAC Address Randomization through Timing
  Attacks}}. In \bibinfo{booktitle}{\emph{ACM Conference on Security and
  Privacy in Wireless and Mobile Networks (WiSec)}}.
\newblock


\bibitem[Minkus and Ross(2014)]%
        {minkus2014know}
\bibfield{author}{\bibinfo{person}{Tehila Minkus} {and}
  \bibinfo{person}{Keith~W Ross}.} \bibinfo{year}{2014}\natexlab{}.
\newblock \showarticletitle{{I Know What You’re Buying: Privacy Breaches on
  eBay}}. In \bibinfo{booktitle}{\emph{Privacy Enhancing Technologies Symposium
  (PETS)}}.
\newblock


\bibitem[Mohtadi and Rahimi(2015)]%
        {mohtadi2015new}
\bibfield{author}{\bibinfo{person}{Hamed Mohtadi} {and}
  \bibinfo{person}{Alireza Rahimi}.} \bibinfo{year}{2015}\natexlab{}.
\newblock \showarticletitle{{New Attacks on Wi-Fi Protected Setup}}.
\newblock \bibinfo{journal}{\emph{{Advances in Computer Science}}}
  (\bibinfo{year}{2015}).
\newblock


\bibitem[of~Transportation: ArcGIS~Online(2025)]%
        {millocations}
\bibfield{author}{\bibinfo{person}{U.S.~Department of Transportation:
  ArcGIS~Online}.} \bibinfo{year}{2025}\natexlab{}.
\newblock \showarticletitle{{Military Bases}}.
\newblock
\newblock
\shownote{\url{https://geodata.bts.gov/datasets/usdot::military-bases/explore}}.


\bibitem[Roberts et~al\mbox{.}(2023)]%
        {roberts2023blue}
\bibfield{author}{\bibinfo{person}{Richard Roberts}, \bibinfo{person}{Julio
  Poveda}, \bibinfo{person}{Raley Roberts}, {and} \bibinfo{person}{Dave
  Levin}.} \bibinfo{year}{2023}\natexlab{}.
\newblock \showarticletitle{{Blue Is the New Black (Market): Privacy Leaks and
  Re-Victimization from Police-Auctioned Cellphones}}. In
  \bibinfo{booktitle}{\emph{IEEE Symposium on Security and Privacy}}.
\newblock


\bibitem[Rye and Levin(2024)]%
        {rye2024surveilling}
\bibfield{author}{\bibinfo{person}{Erik Rye} {and} \bibinfo{person}{Dave
  Levin}.} \bibinfo{year}{2024}\natexlab{}.
\newblock \showarticletitle{{Surveilling the Masses with Wi-Fi-Based
  Positioning Systems}}.
\newblock \bibinfo{journal}{\emph{IEEE Symposium on Security and Privacy}}
  (\bibinfo{year}{2024}).
\newblock


\bibitem[Rye and Beverly(2023)]%
        {ipvseeyou}
\bibfield{author}{\bibinfo{person}{Erik~C Rye} {and} \bibinfo{person}{Robert
  Beverly}.} \bibinfo{year}{2023}\natexlab{}.
\newblock \showarticletitle{{IPvSeeYou: Exploiting Leaked Identifiers in IPv6
  for Street-Level Geolocation}}. In \bibinfo{booktitle}{\emph{IEEE Symposium
  on Security and Privacy}}.
\newblock


\bibitem[Sanatinia et~al\mbox{.}(2013)]%
        {sanatinia2013wireless}
\bibfield{author}{\bibinfo{person}{Amirali Sanatinia}, \bibinfo{person}{Sashank
  Narain}, {and} \bibinfo{person}{Guevara Noubir}.}
  \bibinfo{year}{2013}\natexlab{}.
\newblock \showarticletitle{{Wireless Spreading of WiFi APs Infections Using
  WPS Flaws: An Epidemiological and Experimental Study}}. In
  \bibinfo{booktitle}{\emph{IEEE Conference on Communications and Network
  Security (CNS)}}.
\newblock


\bibitem[Uras et~al\mbox{.}(2022)]%
        {uras2022mac}
\bibfield{author}{\bibinfo{person}{Marco Uras}, \bibinfo{person}{Enrico
  Ferrara}, \bibinfo{person}{Raimondo Cossu}, \bibinfo{person}{Antonio Liotta},
  {and} \bibinfo{person}{Luigi Atzori}.} \bibinfo{year}{2022}\natexlab{}.
\newblock \showarticletitle{{MAC Address De-Randomization for WiFi Device
  Counting: Combining Temporal-and Content-Based Fingerprints}}.
\newblock \bibinfo{journal}{\emph{Computer Networks}} (\bibinfo{year}{2022}).
\newblock


\bibitem[Vanhoef et~al\mbox{.}(2016)]%
        {vanhoef2016mac}
\bibfield{author}{\bibinfo{person}{Mathy Vanhoef},
  \bibinfo{person}{C{\'e}lestin Matte}, \bibinfo{person}{Mathieu Cunche},
  \bibinfo{person}{Leonardo~S. Cardoso}, {and} \bibinfo{person}{Frank
  Piessens}.} \bibinfo{year}{2016}\natexlab{}.
\newblock \showarticletitle{{Why MAC Address Randomization is not Enough: An
  Analysis of Wi-Fi Network Discovery Mechanisms}}. In
  \bibinfo{booktitle}{\emph{Asia Conference on Computer and Communications
  Security (ASIACCS))}}.
\newblock


\bibitem[Viehb{\"o}ck(2011)]%
        {wps-brute}
\bibfield{author}{\bibinfo{person}{Stefan Viehb{\"o}ck}.}
  \bibinfo{year}{2011}\natexlab{}.
\newblock \bibinfo{title}{{Brute Forcing Wi-Fi Protected Setup}}.
\newblock
\newblock
\newblock
\shownote{\url{https://sec-consult.com/fileadmin/user_upload/sec-consult/Dynamisch/Blogartikel/Old_Blogposts/sec-consult-kcodes-netsub-viehboeck.pdf}}.


\bibitem[WiGLE(2025)]%
        {wigle}
\bibfield{author}{\bibinfo{person}{WiGLE}.} \bibinfo{year}{2025}\natexlab{}.
\newblock \bibinfo{title}{{WiGLE -- All the Networks. Found by Everyone.}}
\newblock
\newblock
\newblock
\shownote{\url{https://wigle.net}}.


\bibitem[Wolf et~al\mbox{.}(2020)]%
        {wolf-etal-2020-transformers}
\bibfield{author}{\bibinfo{person}{Thomas Wolf}, \bibinfo{person}{Lysandre
  Debut}, \bibinfo{person}{Victor Sanh}, \bibinfo{person}{Julien Chaumond},
  \bibinfo{person}{Clement Delangue}, \bibinfo{person}{Anthony Moi},
  \bibinfo{person}{Pierric Cistac}, \bibinfo{person}{Tim Rault},
  \bibinfo{person}{Rémi Louf}, \bibinfo{person}{Morgan Funtowicz},
  \bibinfo{person}{Joe Davison}, \bibinfo{person}{Sam Shleifer},
  \bibinfo{person}{Patrick von Platen}, \bibinfo{person}{Clara Ma},
  \bibinfo{person}{Yacine Jernite}, \bibinfo{person}{Julien Plu},
  \bibinfo{person}{Canwen Xu}, \bibinfo{person}{Teven~Le Scao},
  \bibinfo{person}{Sylvain Gugger}, \bibinfo{person}{Mariama Drame},
  \bibinfo{person}{Quentin Lhoest}, {and} \bibinfo{person}{Alexander~M. Rush}.}
  \bibinfo{year}{2020}\natexlab{}.
\newblock \showarticletitle{Transformers: State-of-the-Art Natural Language
  Processing}. In \bibinfo{booktitle}{\emph{Proceedings of the 2020 Conference
  on Empirical Methods in Natural Language Processing: System Demonstrations}}.
  \bibinfo{publisher}{Association for Computational Linguistics},
  \bibinfo{address}{Online}, \bibinfo{pages}{38--45}.
\newblock
\urldef\tempurl%
\url{https://www.aclweb.org/anthology/2020.emnlp-demos.6}
\showURL{%
\tempurl}


\end{thebibliography}
\bibliographystyle{ACM-Reference-Format}

\begin{acronym}
  \acro{AP}{Access Point}
  \acro{AS}{Autonomous System}
  \acro{ASN}{\ac{AS} Number}
  \acro{BGP}{Border Gateway Protocol}
  \acro{BSS}{Basic Service Set}
  \acro{BSSID}{Basic Service Set Identifier}
  \acro{CPE}{Customer Premises Equipment}
  \acro{DAD}{Duplicate Address Detection}
  \acro{EUI}{Extended Unique Identifier}
  \acro{ISP}{Internet Service Provider}
  \acro{IID}{Interface Identifier}
  \acro{LAN}{Local Area Network}
  \acro{mDNS}{Multicast DNS}
  \acro{MAC}{Media Access Control}
  \acro{NIC}{Network Interface Card}
  \acro{NTP}{Network Time Protocol}
  \acro{OCR}{Optical Character Recognition}
  \acro{OS}{Operating System}
  \acro{OUI}{Organizationally Unique Identifier}
  \acro{SOHO}{Small Office-Home Office}
  \acro{SSID}{Service Set Identifier}
  \acro{U/L}{Universal/Local}
  \acro{SLAAC}{Stateless Address Autoconfiguration}
  \acro{VPS}{Virtual Private Server}
  \acro{WPS}{Wi-Fi Positioning System}
\end{acronym}

\end{document}